\documentclass[a4paper,12pt,notitlepage]{article}
\usepackage[latin1]{inputenc}
\usepackage[T1]{fontenc}
\usepackage{graphicx}
\usepackage{amsmath}
\usepackage[comma]{natbib}
\usepackage{booktabs}
\usepackage{sectsty}
\usepackage{times}
\usepackage[a4paper,top=25mm,left=27mm,right=25mm,bottom=30mm,footskip=10mm,footnotesep=8mm]{geometry}
\usepackage{amssymb}
\usepackage{eurosym}
\usepackage{hyperref}
\usepackage{rotating}
\usepackage{multirow}
\usepackage{booktabs}
\usepackage{dcolumn}
\usepackage{subfig}
\usepackage{longtable}

\def\BBS1{../../../../../../alter/FS/BBS1/V04}

\newcommand\bs{\begin{table}[htbp] \scriptsize \centering}
\newcommand\es{\end{table}} 
\newcommand\bss{\begin{sidewaystable}[h] \scriptsize \centering} 
\newcommand\ess{\end{sidewaystable}}

\allsectionsfont{\nohang\noindent}  
\sectionfont{\large}
\subsectionfont{\large}
\subsubsectionfont{\normalsize}
\renewcommand{\baselinestretch}{1.5}
\newcommand{\papertitle}{It's \emph{not} the economy, stupid! How social capital and GDP relate to happiness over time}                               
\newcommand{\acknow}{Francesco Sarracino acknowledges the support of the AFR grant (contract PDR-09-075) by the National Research Fund,
Luxembourg cofunded under the Marie Curie Actions of the European Commission (FP7-COFUND). Stefano Bartolini acknowledges CEPS/Instead for financial support.\\
\noindent
The authors wish to thank Donald William for his kind assistance and insightful comments. The authors are thankful also to Luigi Bonatti and Ma\l gorzata Mikucka for her comments on earlier versions of this paper and to three anonymous reviewers whose comments helped improving this work. A revised version of this article will be published on Ecological Economics. The usual disclaimers apply.}  
\begin{document}

\renewcommand{\baselinestretch}{1}
%\title{\textbf{\papertitle{}}\thanks{\acknow{}}}
\title{\textbf{\papertitle{}}  }

%%% SPECIFY AUTHOR NAMES AND AFFILIATIONS HERE %%%%%%%%%%%%%%%%%%%%%%%%%%%%%%%%%%%%%
\author{
	\textbf{Stefano Bartolini\thanks{Stefano Bartolini}}\\
  University of Siena, Italy and CEPS/INSTEAD, Luxembourg\\
  \vspace{3mm} \\
  \textbf{Francesco Sarracino\thanks{Francesco Sarracino} }\\
  STATEC (Luxembourg) and Higher School of Economics (Russia)}

%%%%%%%%%%%%%%%%%%%%%%%%%%%%%%%%%%%%%%%%%%%%%%%%%%%%%%%%%%%%%%%%%%%%%%%%%%%%%%%%%%%

%% SPECIFY DATE HERE %%%%%%%%%%%%%%%%%%%%%%%%%%%%%%%%%%%%%%%%%%%%%%%%%%%%%%%%%%%%%%
%\date{\today}		
\date{21/05/2014}  
%%%%%%%%%%%%%%%%%%%%%%%%%%%%%%%%%%%%%%%%%%%%%%%%%%%%%%%%%%%%%%%%%%%%%%%%%%%%%%%%%%%

\maketitle

\begin{abstract}
What  predicts the evolution over time of subjective well-being? We correlate the trends of subjective well-being with the trends of social capital and/or GDP. We find that in the long  and medium run  social capital largely predicts the trends of subjective well-being in our sample of countries. In the short-term this relationship weakens. Indeed, in the short run, changes in social capital predict a much smaller portion of the changes in subjective well-being than over longer periods. GDP follows a reverse path, thus confirming the Easterlin paradox: in the short run GDP is more positively correlated to   well-being than in the medium-term, while in the long run this correlation vanishes.\\
\vspace{1cm}

  
\end{abstract}

\section{Introduction}\label{intro}

In the early '70s a ``critique of economic growth'' began to emerge based on the thesis that the pace of polluting emissions and of exploitation of local and global natural resources imposed by industrialization was unsustainable in the long-run \citep{mmrb1972}. More or less in the same period, another stream of critique of growth was initiated by economists such as Galbraith, Scitovsky, Hirsh and Hirschmann that began to question the positive association between income and well-being. However, the latter criticism did not penetrate the mainstream of economic theory.

\citet{easterlin1974} grounded the issue of the relationship between economic growth and well-being on the empirical analysis of self-reported data called subjective well-being (SWB) or happiness\footnote{The reliability of SWB measures has been corroborated by experimental evidence from several disciplines. For example, SWB correlates with objective measures of well-being such as the heart rate, blood pressure, duration of Duchenne smiles and neurological tests of brain activity \citep{bo2008, vanReekun}. Moreover, SWB measures are correlated wtih other proxies of SWB \citep{schwarz, wanous, schimmack} and -- more interestingly -- they mirror the judgements about the respondent's happiness provided by acquaintances or clinical experts \citep{schimmack2, kk2006, layard2005}.}. The evidence provided by Easterlin -- that in the long-run, happiness is not significantly influenced by an increase in income -- has received growing attention from the 1990s onward and overall has had a greater impact in challenging the economic-policy paradigm, which has traditionally emphasized income as one of the principal contributors to human well-being. Although the environmentalist critique of growth still remains the most popular and influential, this second critique also contributed to the revision of national statistics that is currently involving a growing number of National Statistical Offices. 

The lack of correlation over time between average income and average happiness -- labelled the Easterlin paradox -- has been explained by the so-called ``hedonic treadmill'' and ``positional treadmill'' theories. In particular, economists have explored the possibility that these treadmills drive the dynamics of income aspirations, which, in turn, may offset the positive effect of rising income (e.g., \citet{stutzer2004}). The basic idea is that subjective well-being is negatively affected by the level of one's income aspirations. Aspirations may depend either on the income of one's own reference group or on one's own past income. We refer to the first case as the positional treadmill, following the well-rooted tradition in economics and sociology that emphasizes the role of social comparisons and social status (e.g., \citet{veblen1899, duesenberry1949}). We refer to the second case as the hedonic treadmill, following the insights of adaptation theory (see, e.g., \citet{fl1999} and references therein).

Adaptation theory assumes that changes in living conditions (for example, in economic conditions) have a temporary effect on well-being. Neither rising prosperity nor increased adversity durably affects happiness. As time goes by, people tend to revert to their baseline level of well-being. The same mechanism applies to aggregates, such as nations \citep{blanchflower2008}.

Social comparison theory argues that what matters for an individual's satisfaction is his/her relative position with respect to a selected group of people identified as those whom he/she respects and wants to resemble. These people form what is called a ``reference group''  \citep{fk2004, layard2009, Ditella2, ferrer2005, diener1993}. Therefore, the general improvement in income levels brought about by economic growth can result in a negligible increase in average subjective well-being because relative gains and losses compensate each other. A large number of micro-level studies provide evidence in support of both adaptation and social comparison theories \citep{cfs2008}.

Notice that the essence of the Easterlin paradox is the conflict between cross-sections and time series. Indeed, micro-data show that individuals with a higher income than others report higher levels of SWB, at any given point in time. Moreover, cross-country data show that countries with a higher per-capita GDP report higher levels of SWB  \citep{deaton2008, sw2008, inglehart2009, ea2009, fs2002a}. But what about time series? The latter deserve a special attention since they seem more likely than cross-sections to provide an answer to ``what people [...] want to know [...]: How far is general income growth (beyond income levels already achieved) likely to increase average happiness? This is a question about time series relationships'' \citep[][p. 1]{layard2009}. The lack of a relationship between income and happiness is in time series.  

It is now well documented that the time-series of SWB show a substantial heterogeneity across countries \citep{sw2008, inglehart2009}. We know that in the past few decades SWB has increased in some countries and decreased in others, varying at different paces. For instance,  SWB rose in many Western European countries, whereas it fell slightly in the United States. The Easterlin paradox claims that economic growth does not predict the international variability of the time series of SWB. 

All-in-all, the message conveyed by happiness studies -- consisting of a lack of influence of economic growth on well-being, explained by plausible theories and supported by robust empirical evidence -- has contributed to increase the number of those who think that the use of GDP as an indicator of well-being or progress is on the wane. A growing number of scholars feels that it is time to dedicate to ``something else'' -- at least in part -- some of  the enormous attention and policy efforts that contemporary societies pour into economic growth. Several potential candidates have been put forward to assume the role of this ``something else'': social tolerance, political freedom, religiosity, health, social capital, the environment \citep{inglehart2009, deaton2008, vc2006, atm2008, kksss2004, oecd2013, nef2011, dt2012}.

However, whether additional indicators should complement the use of GDP (this position, for example, was taken by the OECD and the Sarkozy Commission \citep{ssf2009}) or entirely replace it (see, e.g., \citet{layard2005} remains a contentious issue. Yet, this disagreement is confined within a growing consensus that GDP ought to play a more limited role than in the past.

However, new developments challenge the message conveyed by happiness studies.  Recently, the robustness of the Easterlin paradox has been questioned in two papers by \citet{sw2008}  and \citet{ssw2010}. These influential papers use the same approach of Easterlin and collaborators based on bivariate analysis, but reach opposite conclusions. They find that GDP and SWB are positively and significantly related over time. The time horizon is the essence of the disagreement between the two research groups. Stevenson, Wolfers and Sachs's sample includes countries with long and short-time series. According to \citet{ea2009} and \citet{eassz2010}  their results depend on the failure to distinguish between the long and the short run. Indeed, Easterlin and collaborators show that GDP matters for SWB in the short run, but this correlation vanishes in the long term. This result is consistent with previous studies identifying  the tendency for SWB and GDP to vary together during periods of contractions and expansions \citep{dtmco2001}.

In a recent study, \citet{clark2011} (see also \citet{cg2010}) put forward a second point further challenging the traditional message conveyed by happiness studies. Clark emphasizes that the potential alternatives to GDP may suffer from the same adaptation and social comparisons effects that prevent economic growth from having a positive impact on well-being in the long-term. Clark argues that scholars dedicated particular attention to the relationship between income and well-being discovering  that adaptation and income comparisons are relevant to this relationship. In contrast, very little efforts were allocated to understand whether social comparisons and adaptation are relevant for the relationship between subjective well-being and its determinants, except GDP. \citet{clark2011} summarizes the small literature which has investigated this issue by concluding that there is some evidence of social comparisons and/or adaptation with respect to unemployment, marriage, divorce, widowhood, the birth of the first child, layoffs, health, social capital and religion. In some cases, as for social capital, his evidence seems weak. \citet{clark2011} cautions against diverting attention towards ``something else'' beyond GDP before we make sure that this something else is not subject to the same shortcomings and concludes that more research is needed. 

Summarizing, these challenging views cast doubts on the traditional message conveyed by happiness studies. Indeed, if GDP turns out to be a good predictor of the variability of the trends of SWB across countries and if the alternative measures to GDP are subject to adaptation and social comparisons, then all this would suggest the need for great caution in downsizing the role of GDP as an indicator of well-being and progress.

The evidence presented in this paper supports instead the view that the message of happiness economics should not change. Since we find that SWB is much more strongly related to social capital than to GDP in the long and the medium run, this suggests that the centrality of GDP should be reduced and social capital should assume a more prominent role than its current one, at least in those social choices that relate to such time horizons.  Indeed,  social capital, as well as economic growth, can also be the target for policies aimed at protecting and boosting it (\citeauthor{helliwell2011}, 2011; \citeauthor{rhgc2010}, 2010; \citeauthor{bartolini2013}, forthcoming). 

The \citet[p. 41]{oecd2001} gives a definition of social capital (SC), consistent with that of \citet{putnam2000}, as ``networks together with shared norms, values and understandings that facilitate co-operation within or among groups''. Several papers have documented that social capital is strongly correlated with SWB in cross-sections (see the pioneering studies by \citet{helliwell2001, helliwell2006} and  \citet{hp2004}; see also \citet{bs2008}, \citet{bpr2008}, \citet{bbp2008}). \citet{brp2009} provided a causal analysis showing that social capital has a strong effect on SWB, using data from Germany. Moreover, even the positive association between religiosity and SWB may be due to social capital, as suggested by \citet{lp2009}, which find that religious people are more satisfied with their lives because they regularly attend religious service and build social networks within their congregations. 

However, the existence of a cross-sectional correlation does not imply the existence of a correlation over time. After all, the contrast between cross-sections and long-term time-series is the essence of the Easterlin paradox. As correctly pointed out by Clark, this contrast may be replicated by any cross-sectional correlate of SWB beyond income. However, the relationship over time between SWB and its correlates is still a largely unexplored issue. In particular, to the best of our knowledge, the techniques for the correlation of time-series adopted by Stevenson, Easterlin and collaborators have never been applied to any other correlate of well-being except GDP. As far as social capital is concerned, we try to fill this gap. 

We provide evidence on the relationship between SWB and social capital by investigating their correlation in the long, medium and short-term. Using time-series from the WVS/EVS and the ESS, we apply the same bivariate methodology which has been applied to analyze the relationship between SWB and growth \citep{sw2008, ssw2010, ea2009, eassz2010}. We find that the trends of social capital are strong predictors of the trends of SWB in the long and medium run, and that their predictive power sharply weakens in the short-term. 

In addition, we provide new evidence on the relationship between SWB and GDP in the medium and short run. GDP follows a reverse path compared to social capital since its importance increases with the decrease in the length of the time horizon. More precisely, GDP does not matter for SWB in the long run, it begins to be relevant in the medium term and its importance grows in the short term. Our results therefore suggest -- in line with the claim by Easterlin and collaborators -- that it is important to distinguish between different time horizons of time series, since results depend on the time span considered. 

This evidence is compatible with both the notion that income is subject to adaptation and social comparisons and with the idea that, vice-versa, social capital is not subject to the same forces. Although the possible existence of spurious correlations and/or endogeneity issues suggests prudence in interpreting these results, they seem to suggest that the road to durable happiness passes more through social capital than through economic growth. This is true at least in developed countries, which constitute the far greater portion of our sample.

The paper is organized as follows: section \ref{data} presents our data, while section \ref{metodo} discusses the relevant methodological aspects. Section \ref{results} presents our findings and section \ref{conclu} concludes.

\section{Data}\label{data}
%PRESENTAZIONE WVS
The set of countries included in our sample depends on the availability of internationally comparable time series on social capital variables, which are very scarce. Our main sources of information are the integrated World Values Survey - European Values Study (WVS/EVS)\footnote{The five wave WVS data-set together with detailed instructions on how to integrate it with the EVS data-set is freely available on-line. For more details, please refer to: http://www.wvsevsdb.com/wvs/WVSData.jsp. The last wave of the EVS is available at the following web address: http://www.europeanvaluesstudy.eu/evs/data-and-downloads}  and the European Social Survey\footnote{http://www.europeansocialsurvey.org} (ESS) data-bases. When analysing the relationship among the variations of SC, GDP and SWB in the long run, we use the WVS/EVS data-set.  For the medium and short run we adopt the  ESS data-base. 

\subsection{World Values Survey and European Values Study data}
The WVS/EVS data-base offers a large compilation of surveys collected in more than 80 countries representing more than 80\% of the world's population. 

This data-base provides information about economic, social, cultural and political variables, surveying nationally representative samples in each wave. In particular the database has information about ``individual beliefs about politics, the economy, religious, social and ethical topics, personal finances, familial and social relationships, happiness and life satisfaction'' \footnote{\citet[p. 6]{bs2008}}. Data have been collected in six waves (1980 - 84; 1989 - 93; 1994 - 99; 1999 - 2004; 2005 - 2007 and 2008 - 2009) for a total of more than 400,000 observations covering a period of about 30 years. 

The present study focuses on a smaller sample of 27 countries, however, for a total of about 169,000 observations.  This restriction is imposed by the limited availability of long time-series  for our variables of interest. We consider as long-term a time horizon of at least 15 years.

Our sample is further restricted to countries with at least 3 waves of observations for both SWB and SC variables. The reason for this choice is to reduce the risk that the trends of the relevant variables are  affected by wave-specific biases due to shocks and/or measurement errors. This choice is a reasonable compromise between using only two waves -- which would maximize the above mentioned risk -- and using four or more waves, which would excessively reduce our sample size.  

Furthermore,  our sample does not include transition economies because in the first years of the transition to capitalism the economic, cultural and institutional shock was so dramatic that, arguably, it deeply affected SWB far beyond the evolution over time of SC or GDP. The inclusion of the waves collected close to the beginning of the transition, which are presumably strongly influenced by such powerful confounders, might result in misleading conclusions. At the same time, if we were to exclude the observations that were collected close to the institutional shock of 1989, no transition country would satisfy our long time-span requirement. Therefore, the need to monitor the relationship over time of our variables in relatively stable conditions required the exclusion of transition economies from the long-term analysis.

Table~\ref{desc-a008r-01V04}   in  \ref{desctabs} summarizes the cross-country and waves availability of observations for the long run analysis considering two proxies of SWB.

SWB in WVS/EVS is observed through the answers to two questions: the first one regarding the overall feeling of happiness and the second one   about the respondent's satisfaction with life. More specifically, the first variable ranges on a 1 to 4 scale and is based on answers to the question: \emph{``All considered you would say that you are: 1. very happy; 2. pretty happy; 3. not too happy; 4. not at all happy?''}. This variable has been recoded so that the category ``very happy'' corresponds to the highest value in the scale and the category ``not at all happy'' corresponds to the lowest one.

The second measure of SWB is the so-called ``life satisfaction''. This variable is observed through the question: ``all things considered, how satisfied are you with your life as a whole these days?''. Possible answers range  on a 1 to 10 scale in which the lowest value corresponds to ``dissatisfied'' and the highest to ``satisfied''. 

The two proxies of SWB are not always observed in the same wave. Thus, our analyses are based on sub-sets of the data when each of the two proxies of SWB were observed jointly with the proxies of SC. Conversely,  availability of data about GDP do not raise any problem. 

We proxy individual SC by observing the respondent's participation in various kinds of groups and associations. During interviews, people are asked whether they are members or not of a list of  groups or associations. This list is quite large and contains participation in religious, cultural, sport, professional and many other kind of associations (for the complete list of groups or associations see   \ref{list}). We created a dichotomous variable taking values of 1 if the respondent declares to participate in at least one group or association, 0 otherwise. 

Finally, we include data about GDP per capita (constant 2000 US\$) from the World Development Indicators (WDI).\footnote{World Development Indicators and Global Development Finance, http://databank.worldbank.org/ddp/home.do?Step=12\&id=4\&CNO=2 We excluded Northern Ireland from our sample because the World Development Indicators provide no data on GDP for this country.}
Consistent with previous studies, we used the logarithm of GDP per capita to take into account the non-linear relationship between subjective well-being and GDP \citep{eassz2010, ssw2010}.

Descriptive data and missing values for each variable are presented in table \ref{miss-a008r}  and table \ref{miss-a170r} in  \ref{missWVS}  for happiness and life satisfaction data, respectively. The small number of missing data and the absence of specific patterns of missingness rule out the risk of biased estimates \footnote{Please refer to  \ref{missWVS}  for a detailed description of available data.}

\subsection{European Social Survey data}
When computing the short and the medium-term variations we resort to the ESS. To shorten the time horizon of our analysis we could in principle split our period of observation in the WVS/EVS into shorter sub-periods defined by a certain distance between the waves. However, in our WVS/EVS sample the distances between two consecutive observations are highly irregular, ranging from 1 to 14 years. As a consequence, in the WVS/EVS it is not possible to attribute the variations between contiguous waves to the long, medium or short run.  Table \ref{table-a008r}  in  \ref{desctabs} provides an overview of the distances among waves in all the considered countries\footnote{The table about life satisfaction has been omitted for brevity, but it is available on request to the authors.}. Having data that are measured at regular intervals is the key point to identify which time horizon is measured by such intervals.

Therefore, when computing the medium and short-term changes we use ESS data whose time-series mostly reach 6 years, a reasonable medium-term. Moreover, it is possible to compute short-term changes by splitting the 6 year period of observations into the shortest possible sub-periods, defined by the interval between contiguous waves. In the ESS this interval is two years for almost all countries. Therefore, we split the available ESS time series into biannual intervals for each country and we compute the variation from one wave to another for each variable separately. Biannual intervals are short enough to be considered as short-term.

However, this choice has a cost: the ESS observes associational activities only in two years (2002 and 2004) -- a  time too short for our purposes -- whereas it provides time-series about social trust covering the whole period. This forces us to adopt social trust as a proxy of social capital in the ESS. This shift in the measure of social capital makes the comparability of the results from the long-term with those from shorter time periods questionable. To provide some evidence concerning such comparability we check the consistency of the estimates from the medium-term of the ESS with those of the WVS/EVS. Medium-run estimates are possible in the WVS/EVS by keeping in the sample only those waves whose distance is comprised between 3 and 6 years. However, this is an approximate test   because if the WVS/EVS were observed at regular intervals, for instance every 5 years, the number of observations would be approximately tripled. Actually instead, the number of observations increases marginally when moving from the long to the medium-term, because the number of countries drops (from 27 to 19). Hence, the sample of countries available for the medium-term analysis in the WVS/EVS somewhat differs from the one used for the long-term analysis. 

The European Social Survey  was first run in 2002 and, since then, it has been conducted regularly every two years in 2004, 2006 and 2008. The ESS is designed to observe the interaction between institutions and  people's attitudes, beliefs and behaviours across Europe.  This feature makes ESS a useful source of data for the present study since it provides, among others, information about SC and well-being on a relatively large sample of countries surveyed at regular intervals over time. However,  given its European perspective, it provides information on a smaller number of countries (about 30) than the WVS/EVS. 

Table \ref{desccountries_happy}   in  \ref{desctabs} summarizes the cross-country and wave availability of observations for the shorter run analysis and for the two available proxies of SWB respectively.

The actual sample size includes 24 countries for a total of about 153,800 observations.  It is  constituted by western European countries, transition economies from Eastern Europe, Israel and Turkey. In this case, we included transition countries in our ESS sample because they started being surveyed more than 10 years after the institutional shock. Arguably, such a period is long enough to make the impact of the collapse of socialism on our variables negligible.  Finally, Bulgaria, Cyprus, Italy, Luxembourg and the Russian federation have been excluded because they have been observed only in 2002 and 2004, a  period too short to perform a medium-term analysis.

As with the WVS/EVS, the ESS questionnaire also includes questions on happiness and life satisfaction. The wording of the life satisfaction question is identical to the one asked in the WVS/EVS. The only difference is that the answers are on an eleven point scale instead of ten (0 means extremely dissatisfied and 10 means extremely satisfied). 

The wording of the happiness question is only slightly different from the WVS/EVS  (``taking all things together, how happy would you say you are?'', while the possible answers range from 0 (``extremely unhappy'') to 10 (``extremely happy''), instead of the four point Likert scale of the WVS/EVS. 

As said, in the ESS the only measure of social capital available for the whole period is social trust.  More precisely, trust is proxied by answers to three questions. Respondents have been asked to rate their perceptions about whether most people can be trusted or not, whether other people  try to take advantage of them and whether they try to be helpful or rather looking for themselves. Each of these three questions ranges on a 0 to 10 scale, where the lowest category corresponds to the worst judgement and the highest to the best one.\footnote{Notice that the WVS/EVS provides long time series on a trust question. The respondents are asked whether most people can be trusted or not with a wording very similar to the ESS analogous question. However, different from the ESS, the answer is dichotomous in the WVS/EVS (yes/no). Arguably, the answer on an eleven point scale in the ESS provides a better and more sensible scaling for the answer than the  binary one from the WVS/EVS.}

Given the similarities among these three questions, both in terms of wording and in terms of substantive meaning, we run a factor analysis to check whether they could be grouped to proxy one latent concept. We first performed a factor analysis on the pooled sample (see tab.\ref{factor_tot}) and subsequently we analyzed the sample wave by wave (see tab.\ref{factor}). In both cases factor loadings suggest that the three variables are largely mirroring the same fundamental concept that we label social trust. Therefore, in our regressions we use the social trust index as obtained by means of factor analysis\footnote{For more details please refer to  \ref{facapp}}.

Finally,  we use the logarithmic form of the GDP per capita (constant 2000 US\$) from the World Development Indicators (WDI)\footnote{World Development Indicators and Global Development Finance, http://databank.worldbank.org/ddp/home.do?Step=12\&id=4\&CNO=2}.

Table \ref{miss-ESS} and tab.\ref{tabmiss-ESS} in  \ref{missessapp} report descriptive statistics and percentages of missing data for the considered variables. The percentages of missing data are small enough to rule out the risk of biased estimates.

\section{Empirical strategy}\label{metodo}

Previous empirical work about the relationship between economic growth and SWB over time are based on bivariate regressions of aggregate measures of SWB and per capita income \citep{sw2008, ssw2010, ea2009, eassz2010}. Since our primary focus is to investigate the relationship between SC and SWB over time, a natural strategy is to adopt the same bivariate approach, where of course we substitute SC for GDP  in our baseline regression model (see eq. \ref{eq4} and eq. \ref{eq5}). Moreover, we also aim at comparing the potential of the time series of social capital and GDP to predict the variation over time of SWB. 

To these aims, we develop our empirical strategy in three steps: i) we compute the trends of the  proxies of SC, GDP and SWB; ii) we run bivariate regressions of the trends of SWB on the trends of SC or log of GDP per-capita, separately. The second specification is basically meant to replicate on our samples what has been done by previous studies on the relationship between SWB and GDP over time; iii)  we provide trivariate regressions of SWB on both trends of log GDP and SC to account for eventual spurious correlations. 

The risk of spurious correlations should not be underestimated. Indeed, the literature on economic growth and SC pointed out that these two variables may be related to each other in many ways \citep{KK1997, roth2009, zk2001}. For instance, \citet{pln1993} showed that there are paths through which SC fosters economic growth. Conversely, there is also a long standing tradition emphasizing that economic growth can erode the stock of SC over time \citep{polanyi1968, hirsch1976} (see also \citet{bb2008}). An implication of the possible relationships between GDP and SC is that the bivariate correlations with SWB might be affected by spurious correlations. However, our findings from trivariate analysis seem to rule out this possibility.

\subsection{Estimating trends}\label{methodsub1}
We compute the long and the medium-term trends for the various proxies of SC and SWB by regressing them on a time variable containing all the years when the dependent variable was observed \citep{ea2009, eassz2010}. Trends are computed for each country separately. The coefficient of the time variable represents the estimated average yearly variation for the specific dependent variable. 

Since we have various indicators of SC and  SWB, our regression methodology changes depending on the nature of the dependent variable: in case of a dichotomous variable (i.e. membership in groups or organizations), we adopted a probit model with robust standard errors reporting marginal effects. 
The resulting equation is:

\begin{equation}
	Pr(Proxy_{ij} = 1 | YEAR_{ij}) = \phi(\beta_{j} \cdot YEAR_{ij} + \mu_{ij})
	\label{eq1}
\end{equation} 
\noindent
where $\phi$ is a normal cumulative distribution function.  Index $j$ stands for the various proxies of SC and SWB, while index $i$ stands for individuals. Marginal effects of coefficients are subsequently computed.

In the case of an ordered dependent variable taking discrete values (i.e. feeling of happiness or satisfaction with life) ordered probit or logit models should be applied \citep{ferrer2005}. However, there is robust evidence that in such cases the use of an OLS model is equivalent to these alternative techniques in terms of the sign and of the significance of the coefficients \citep{fcf2004, blanchflower2008}. Moreover, OLS models have a strong advantage: they allow a direct comparison between regressors from various regressions. Therefore, we adopt the following OLS model:

\begin{equation}
	Proxy_{ij} = \alpha + \beta_{j} \cdot YEAR_{ij} + \mu_{ij}
	\label{eq2}
\end{equation} 

The same equation is also adopted to compute the trend of the index of social trust (in the ESS) and of the logarithm of GDP per capita. 

Previous work uses different methods to compute economic growth. \citet{ea2009} and \citet{eassz2010} used the growth rate of the logarithm of GDP, while \citet{sw2008} and \citet{ssw2010} adopted the difference between the logarithm of GDP at the beginning and at the end of the period. Both specifications overlook what happened to GDP between the initial and the final year of the time series. The problem with this choice is that it ignores the intermediate information, thereby increasing the risk that the variation of GDP is affected by wave-specific biases due to shocks and/or measurement errors. Our estimation of the yearly variation of the logarithm of GDP reduces this risk because it uses also the intermediate information. 

To compute the short-term variations we split our period of observation into the shortest possible sub-periods, defined by the interval between contiguous waves (see section \ref{data}). This exercise is possible only with ESS data where such intervals are regular. In this case we simply calculate the variation of the variable of interest in the interval defined by two consecutive waves.% Therefore, we compute the short term variations in a sensibly different way from the one applied to compute the long and medium run trends. In the first case it is indeed a variation between two consecutive waves, while in the second one it is the estimated trend over the whole period.

Long, medium and short-run changes in our variables have been computed applying the original weights provided in WVS/EVS or ESS.

\subsection{Bivariate and trivariate analysis}\label{methodsub2}
To check the correlation among respectively long, medium and short-run changes of SWB and SC or GDP we run bivariate linear regressions with robust standard errors. Formally, we estimate the two following models:

\begin{equation}
SWB^{trend}_{j} = \alpha + \beta \cdot SC^{trend}_j + \mu_j 
\label{eq4}
\end{equation}

\begin{equation}
SWB^{trend}_{j} = \alpha + \beta \cdot lnGDP^{trend}_j + \mu_j 
\label{eq5}
\end{equation}
\noindent
where $SWB^{trend}$, $SC^{trend}$ and $lnGDP^{trend}$ represent   estimated variations of SWB, SC and GDP as previously computed; $\mu$ is the error term and the index $j$ refers to countries. To allow comparability of coefficients within regression models, we used standardized variables.  The variables are standardized by first subtracting their average  and subsequently dividing by their standard deviation.

Notice that our method is different from the one applied by Easterlin and colleagues. They measure short-run variations of SWB and the logarithm of GDP as the ``deviation at each date of the actual value from the trend value''\footnote{\citet[p. 3]{eassz2010}}, thus defining the short term as a departure from the long term trend. Differently from Easterlin and colleagues, our method allows us to directly compare short-term coefficients with longer term ones from eq. \ref{eq4}, \ref{eq5} and \ref{eq6}. 

To check the possibility that our bivariate regressions are the outcome of spurious correlations we also run a set of trivariate regressions in which we correlate the variations of SWB with the variations of both SC and GDP. Hence, we test a linear model with robust standard errors resulting in the following trivariate equation:
\begin{equation}
SWB^{trend}_{j} = \alpha + \beta_{1} \cdot  SC^{trend}_j + \beta_{2} \cdot  lnGDP^{trend}_j + \mu_j 
\label{eq6}
\end{equation}
\noindent
where the only difference with eq. \ref{eq4} is that a third term including the change of the logarithm of GDP has been added.

\section{Results}\label{results}
\subsection{The long-term (15 years)}
\begin{figure}[htbp]
\centering
\subfloat[][\emph{feeling of happiness.\label{h_yindex23}}]
{\includegraphics[width=.65\columnwidth]{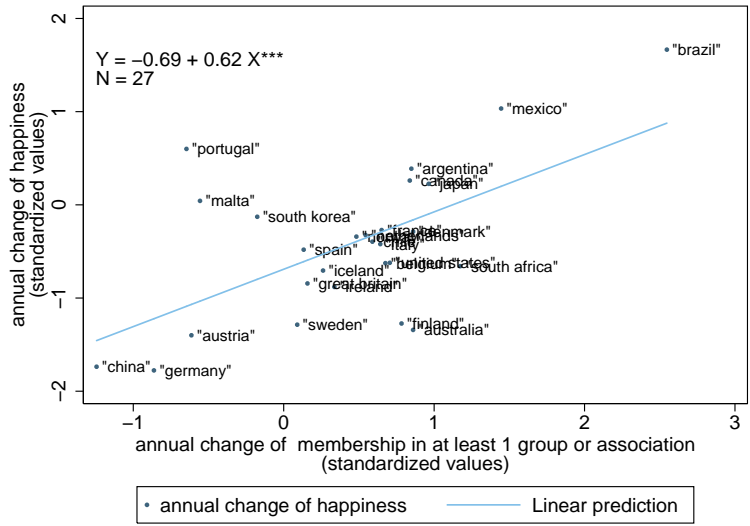}} \quad
\subfloat[][\emph{satisfaction with life.\label{ls_yindex23}}]
{\includegraphics[width=.65\columnwidth]{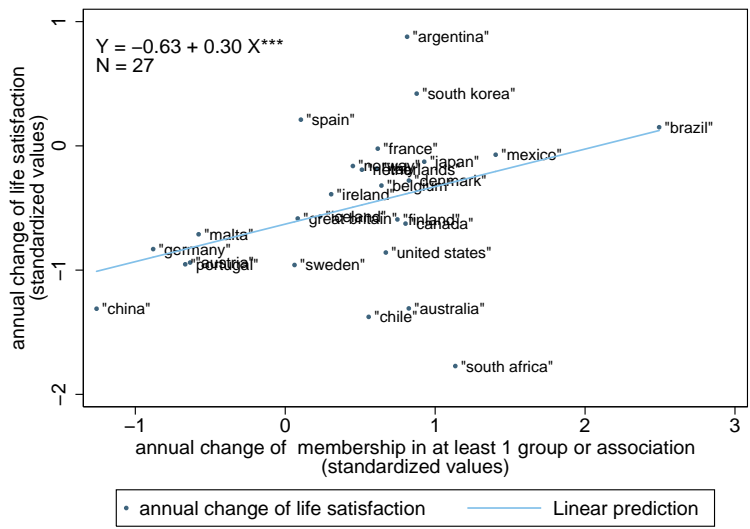}} \\
\caption{Correlations among long-term trends of proxies of subjective well-being and of social capital. Each dot on the scatterplot associates the long-term trend of SWB - on the y axis - with the long-term trend of group membership for each country. The regression line simply depicts the correlation between the two variables.}
\label{lp-03-happy}
\end{figure}

In the long run  changes in both life satisfaction or happiness are strongly and positively correlated to the trends of SC. Figures \ref{h_yindex23} and \ref{ls_yindex23} graphically summarize this result. An increase by one standard deviation in the trend of group membership  is associated with a 0.62 point increase in the trend of happiness and 0.30 point increase in the trend of life satisfaction.

\begin{figure}[htbp]
\centering
\subfloat[][\emph{Happiness and the the logarithm of GDP.\label{h_yindex22}}]
{\includegraphics[width=.65\columnwidth]{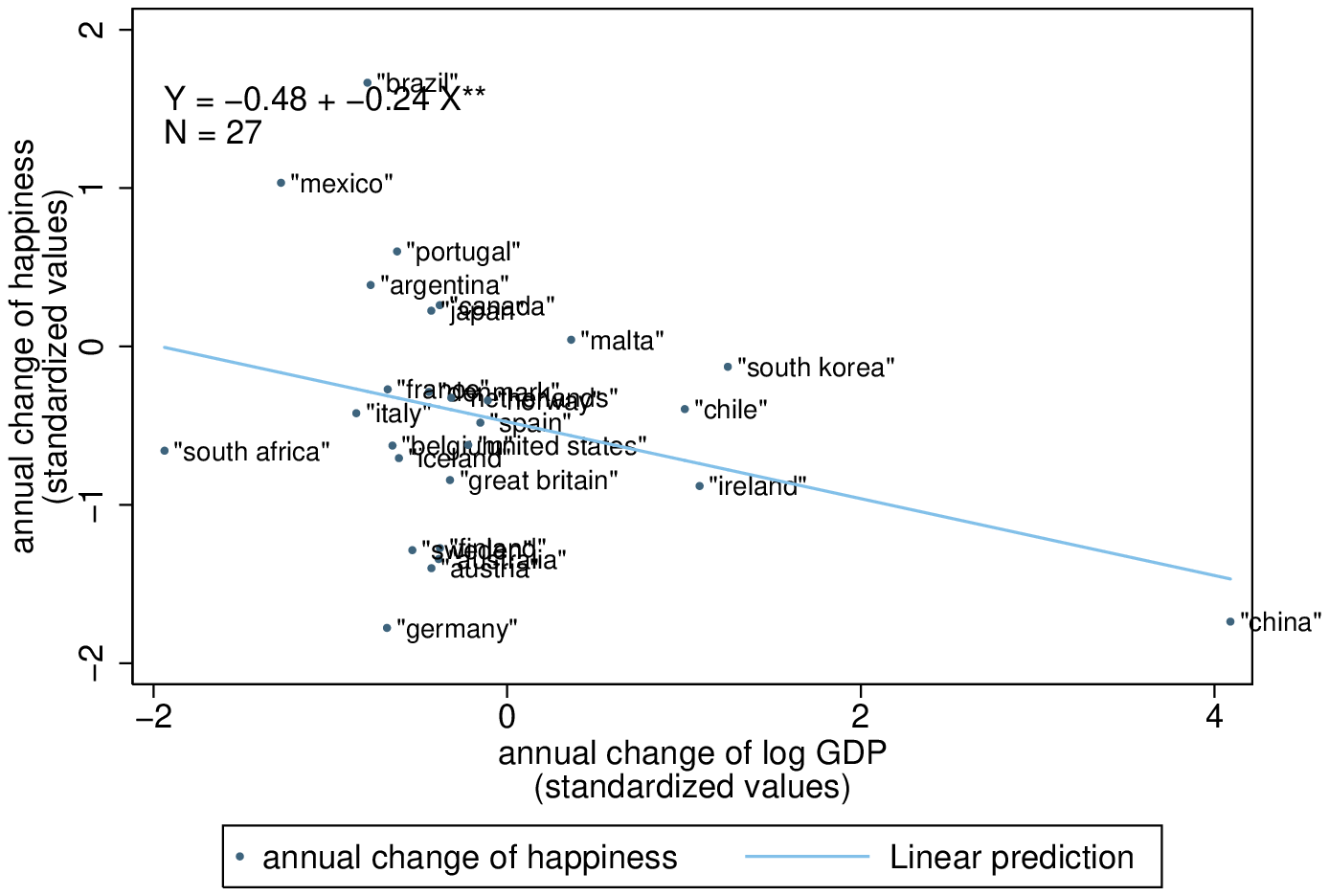}} \quad
\subfloat[][\emph{Life satisfaction and the logarithm of GDP.\label{ls_yindex22}}]
{\includegraphics[width=.65\columnwidth]{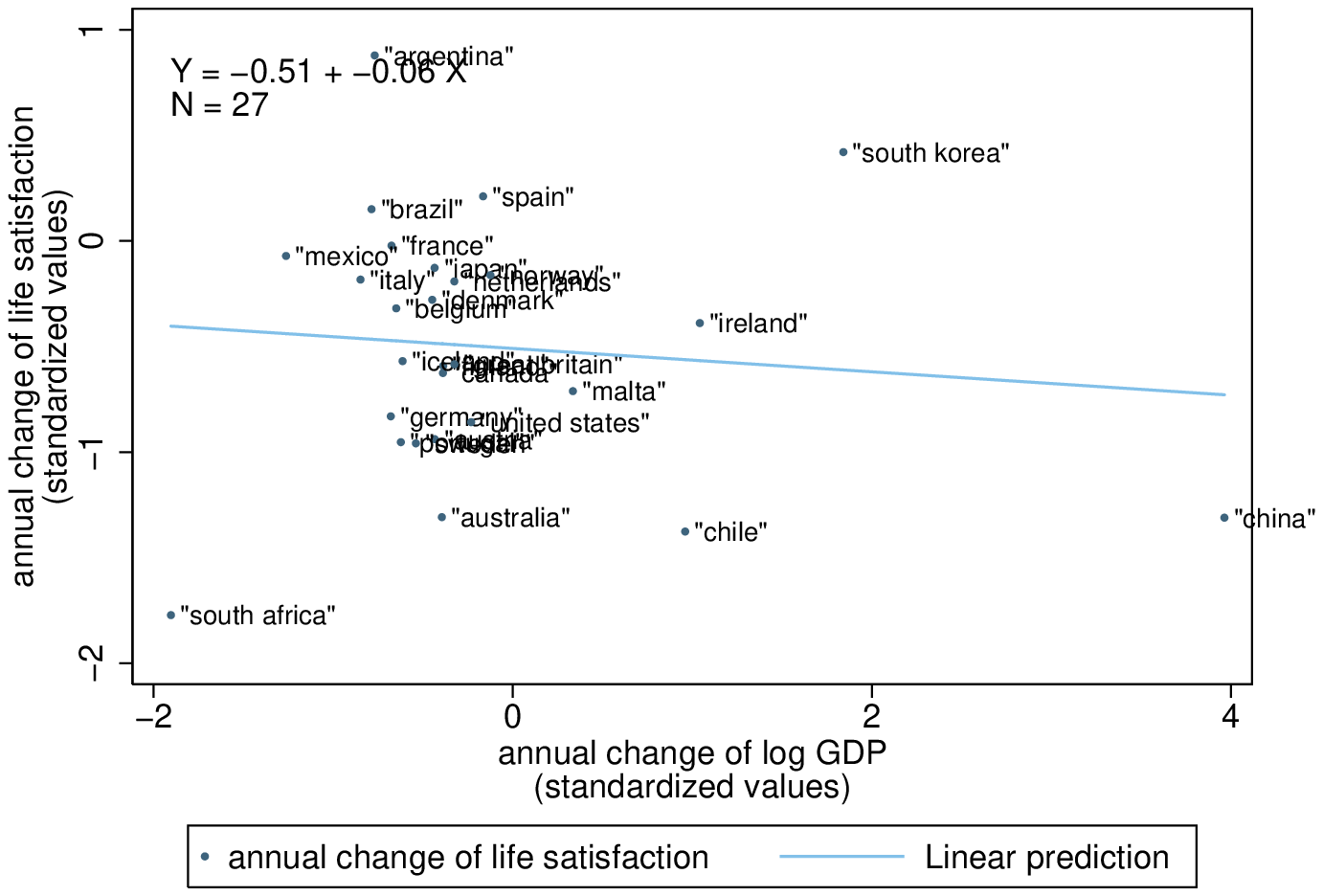}} \\
\caption{Correlations among long-term trends of proxies of subjective well-being and of the logarithm of GDP. Each dot on the scatterplot associates the long-term trend of SWB - on the y axis - with the long-term trend of GDP. The regression line simply depicts the correlation between the two variables.}
\label{lp-03-happy}
\end{figure}

Figures \ref{h_yindex22} and \ref{ls_yindex22} inform that when we substitute GDP for social capital, its long run trends are unrelated to the trends of life satisfaction and negatively and significantly correlated with the trends of happiness. 

Results from trivariate analysis substantially confirm the evidence from bivariate analysis (see tab. \ref{lp_03_trivar_swb}). The dimension and the significance of the coefficients of SC remains very similar to the ones resulting from bivariate analysis. The only exception concerns the long run correlation between happiness and GDP, which turns out to be non-significant.

\begin{table}[htbp]\centering
\def\sym#1{\ifmmode^{#1}\else\(^{#1}\)\fi}
\caption{Trivariate regressions of long-term trends of proxies of subjective well-being over trends of SC and GDP (standardized variables).}
\label{lp_03_trivar_swb}
\begin{tabular}{l*{2}{D{.}{.}{-1}}}
\toprule
                &\multicolumn{1}{c}{(1)}&\multicolumn{1}{c}{(2)}\\
                &\multicolumn{1}{c}{happiness}&\multicolumn{1}{c}{life satisfaction}\\
\midrule
membership in group or association&    0.608\sym{**} &    0.330\sym{**} \\
                &   (2.19)         &   (3.58)         \\
\addlinespace
log GDP         &  -0.0100         &   0.0447         \\
                &  (-0.07)         &   (0.35)         \\
\addlinespace
Constant        &   -0.690\sym{***}&   -0.634\sym{***}\\
                &  (-3.88)         &  (-6.87)         \\
\midrule
Observations    &       27         &       27         \\
Adjusted \(R^{2}\)&    0.302         &    0.087         \\
\bottomrule
\multicolumn{3}{l}{\footnotesize \textit{t} statistics in parentheses}\\
\multicolumn{3}{l}{\footnotesize \sym{*} \(p<0.10\), \sym{**} \(p<0.05\), \sym{***} \(p<0.001\)}\\
\end{tabular}
\end{table}

\citet{sw2008} claim that in some waves of the WVS/EVS the samples of some countries are not representative of the overall population\footnote{The list includes the first three waves of Argentina, Chile, China and India and the first wave of South Africa. After excluding the first three waves, the first four countries do not satisfy the requirement of 15 years length of time series. Hence, they are exluded from the sample in our robustness check. Please, refer to   \ref{probsampling} for more details.}. However, as reported in  \ref{probsampling}  our results are robust to the exclusion of those countries. 

The message of our long-term analysis is that SC matters a lot in predicting the trends of SWB. As far as GDP is concerned, our results are consistent with the Easterlin paradox.

\clearpage
\subsection{The medium-term (3-6 years)}
\citet{ea2009} and \citet{eassz2010} argue that the relationship between SWB and GDP changes if we shift the focus from the long run to shorter periods.

Do the results on the relationship between SWB and SC also vary when we consider a shorter time horizon? We try to answer this question by shifting our analysis from the long to the medium and  short run. For the reasons described in section \ref{data}, we first perform our analysis adopting ESS data and then we turn to WVS/EVS data to provide an approximate test of the comparability of the results using different measures of social capital.

For our medium-term estimates, most of the countries of our sample have at least 6 years of time between the first and the last wave. Austria, Estonia, Slovakia and Ukraine have been surveyed in three out of four waves. In these cases the maximum available time-span is 4 years. 

\begin{figure}[htbp]
\centering
\subfloat[][\emph{Happiness and the index of social trust.\label{lp_hap_yindex23}}]
{\includegraphics[width=.65\columnwidth]{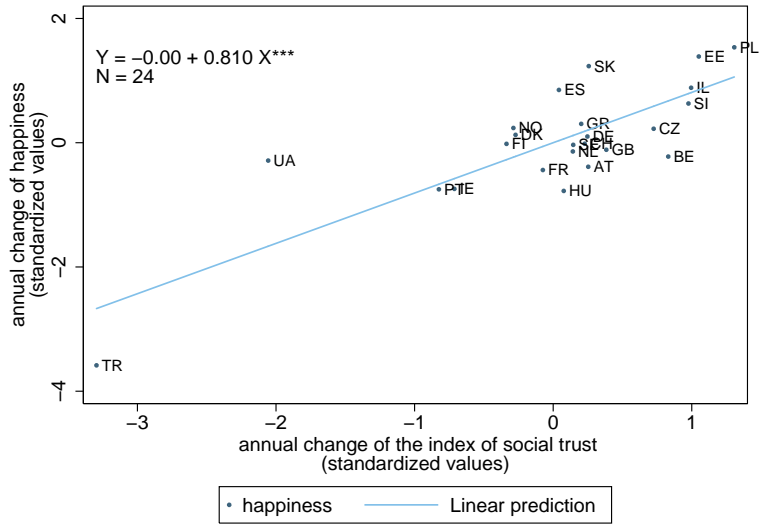}} \quad
\subfloat[][\emph{Life satisfaction and the index of social trust.\label{lp_lsat_yindex24}}]
{\includegraphics[width=.65\columnwidth]{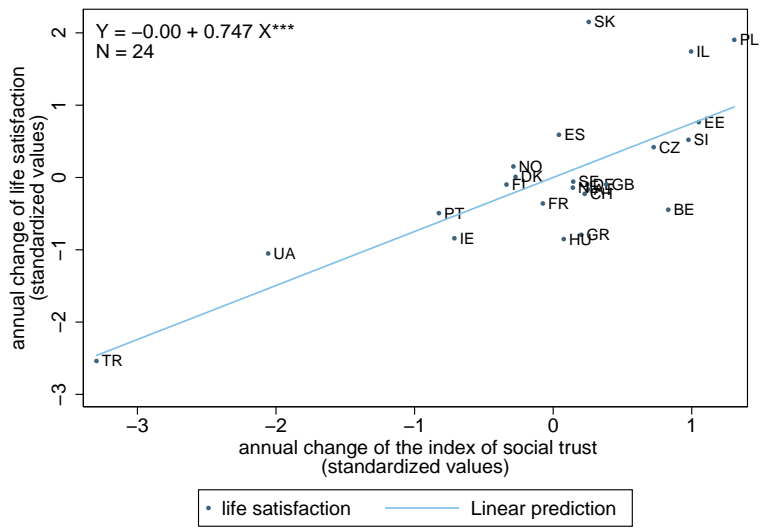}} \\
\caption{Correlations among medium-term trends of subjective well-being and the changes in the index of social capital. Each dot on the scatterplots associates the medium-term trend of SWB - on the y axis - with the medium-term trend of the index of social trust. The regression line simply depicts the correlation between the two variables.}
\label{subfig2}
\end{figure}

\begin{figure}[htbp]
\centering
\subfloat[][\emph{Happiness and the the logarithm of GDP.\label{lp_hap_gdp}}]
{\includegraphics[width=.65\columnwidth]{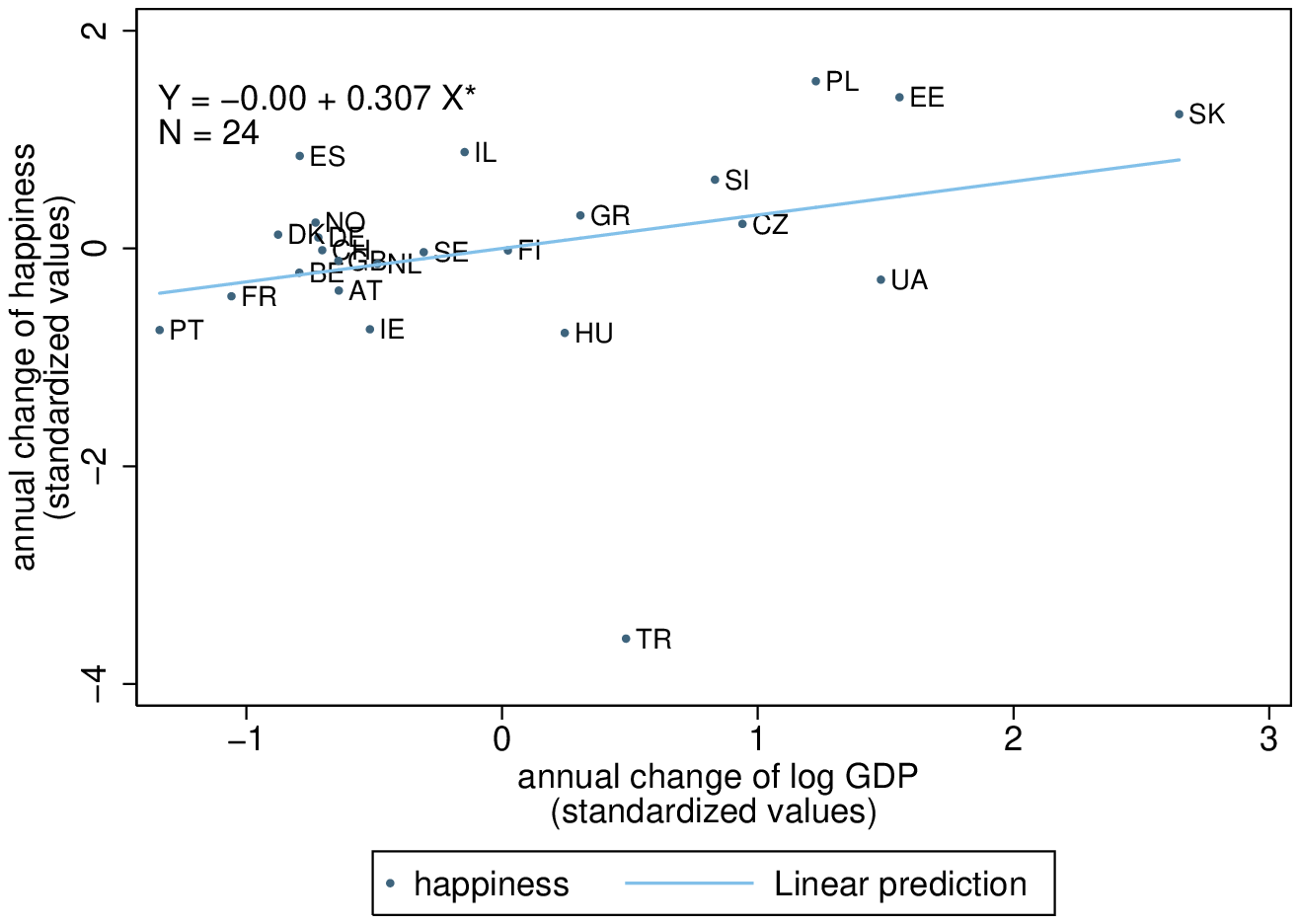}} \quad
\subfloat[][\emph{Life satisfaction and the logarithm of GDP.\label{lp_lsat_gdp}}]
{\includegraphics[width=.65\columnwidth]{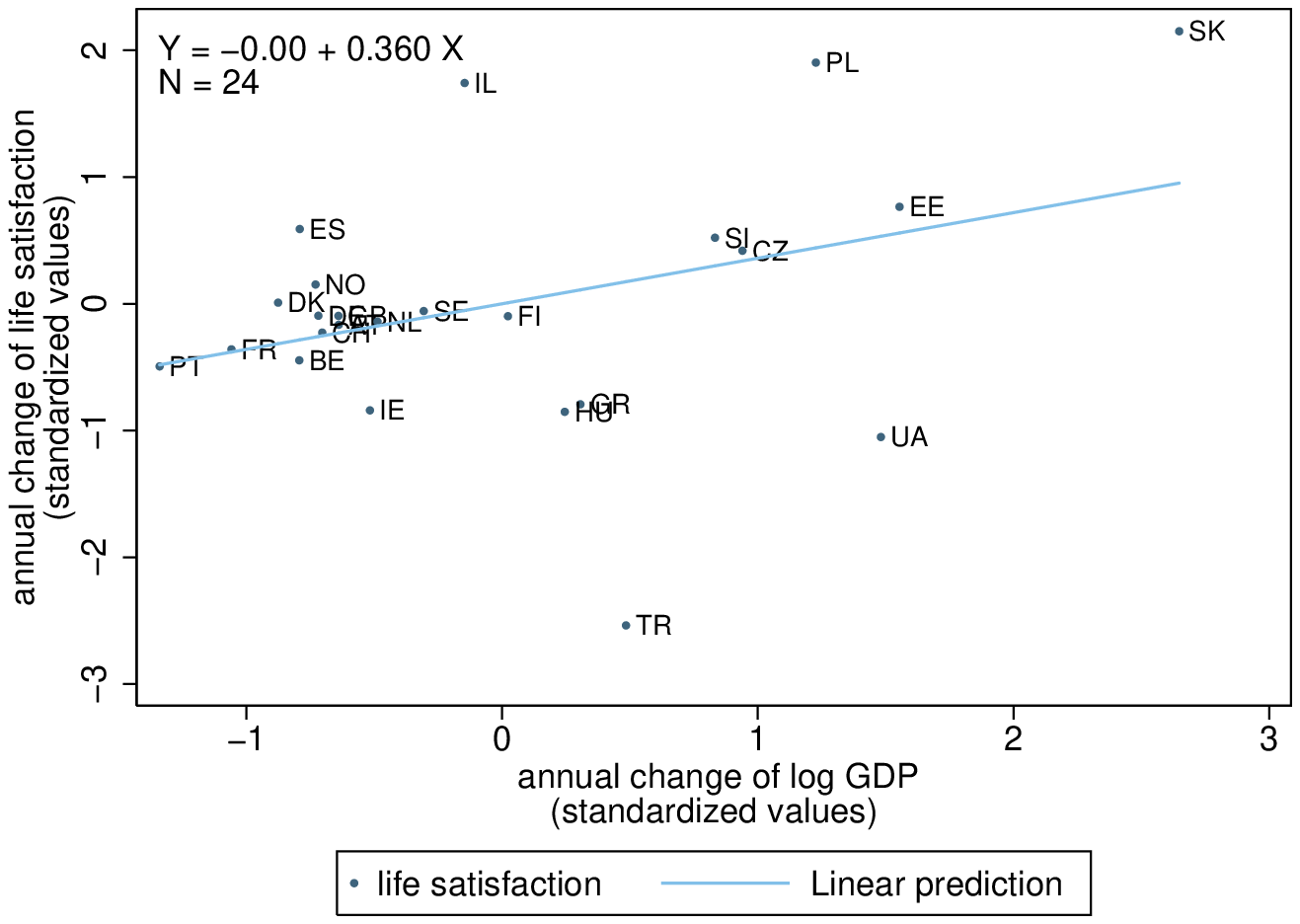}} \\
\caption{Correlations among medium-term trends of subjective well-being and of the logarithm of GDP. Each dot on the scatterplots associates the medium-term trend of SWB - on the y axis - with the medium-term trend of the logarithm of GDP. The regression line simply depicts the correlation between the two variables.}
\label{subfig2}
\end{figure}

In the medium-term we find a positive bivariate correlation between the changes in SWB and in the index of social trust. Figures \ref{lp_hap_yindex23} and \ref{lp_lsat_yindex24} graphically summarize this result\footnote{The list of country acronyms in the ESS data-base is available in  \ref{acronyms}.}. 

Coefficients are large in both happiness and life satisfaction regressions. One standard deviation in the change of the index of social trust correlates with a 0.81 point increase in the variation of happiness and 0.75 point increase for life satisfaction. The position of Turkey in the scatterplot -- looking like an outlier -- can cast  doubt that our result is driven by the inclusion of this country. However, this is not the case. Even if we delete this country, the coefficients are still large and significant \footnote{See tab. \ref{lp_trivar_swb_factor-turkey}  in   \ref{mediumterm}  for a check excluding Turkey from the sample.}.

As for GDP, figures \ref{lp_hap_gdp} and \ref{lp_lsat_gdp} indicate that the coefficient of GDP turns positive and weakly significant for happiness, while it remains non-significant for life satisfaction.

This latter coefficient turns out to be significant (at 10\%) in trivariate regressions while the coefficient for happiness increases its significance (at 5\%). Both coefficients maintain a similar magnitude compared to the bivariate analysis (see tab. \ref{lp_trivar_swb_factor}). 

When considering the variations of SC, trivariate analysis confirms the results of bivariate regressions, both in terms of the large magnitude of the coefficients (0.79 for happiness and 0.73 for life satisfaction) and in terms of their high significance. 

\begin{table}[htbp]\centering
\def\sym#1{\ifmmode^{#1}\else\(^{#1}\)\fi}
\caption{Trivariate regressions of trends of subjective well-being over changes of the index of social trust and trends of GDP (standardized variables).}
\label{lp_trivar_swb_factor}
\begin{tabular}{l*{2}{D{.}{.}{-1}}}
\toprule
%                &\multicolumn{1}{c}{(1)}&\multicolumn{1}{c}{(2)}\\
                &\multicolumn{1}{c}{happiness}&\multicolumn{1}{c}{life satisfaction}\\
\midrule
index of social trust&    0.797\sym{***}&    0.731\sym{***}\\
                &   (4.03)         &   (8.06)         \\
\addlinespace
trend of log GDP&    0.268\sym{**} &    0.323\sym{*}  \\
                &   (2.41)         &   (2.02)         \\
\addlinespace
Constant        &-7.96e-10         & 5.56e-10         \\
                &  (-0.00)         &   (0.00)         \\
\midrule
Observations    &       24         &       24         \\
Adjusted \(R^{2}\)&    0.702         &    0.630         \\
\bottomrule
\multicolumn{3}{l}{\footnotesize \textit{t} statistics in parentheses}\\
\multicolumn{3}{l}{\footnotesize \sym{*} \(p<0.10\), \sym{**} \(p<0.05\), \sym{***} \(p<0.001\)}\\
\end{tabular}
\end{table}

The coefficients of SC turn out to be more than 2 times bigger than the ones of GDP and more statistically significant.  Interestingly, both the magnitude and the significance of the coefficients of SC and of the logarithm of GDP are similar in  both happiness and life satisfaction regressions. 

\begin{table}[htbp]\centering
\def\sym#1{\ifmmode^{#1}\else\(^{#1}\)\fi}
\caption{Trivariate correlations among medium-term trends of feeling of happiness, of social capital and economic growth using WVS/EVS data (standardized values).}
\label{lp_trivar_swb_factor_WVSEVS}
\begin{tabular}{l*{1}{D{.}{.}{-1}}}
\toprule
                &\multicolumn{1}{c}{happiness}\\
\midrule
membership in group or association     &    0.240\sym{**} \\
                &   (2.31)         \\
\addlinespace
log GDP     &    0.231         \\
                &   (1.67)         \\
\addlinespace
Constant        &   -0.361\sym{**} \\
                &  (-2.85)         \\
\midrule
Observations    &       36         \\
Adjusted \(R^{2}\)&    0.153         \\
\bottomrule
\multicolumn{2}{l}{\footnotesize \textit{t} statistics in parentheses}\\
\multicolumn{2}{l}{\footnotesize \sym{*} \(p<0.10\), \sym{**} \(p<0.05\), \sym{***} \(p<0.001\)}\\
\end{tabular}
\end{table}

These results hold also after excluding transition economies and Turkey from the sample (see tab. \ref{lp_trivar_swb_factor-tt} in  \ref{mediumterm}). However, in this case the significance of the coefficients is reduced, but this might be the outcome of the smaller sample size (16 countries).

Moreover, the relationships we identify are basically confirmed when we run the medium-term analysis using associational activity as a measure of social capital and WVS/EVS data (see tab. \ref{lp_trivar_swb_factor_WVSEVS}).  The main difference in this case concerns life satisfaction since in this case the coefficient of social capital is non-significant. However, previous studies have questioned the reliability of life satisfaction in the WVS/EVS  \citep{sw2008}. Instead, figures about happiness are consistent with those from the ESS: the trends of associational activity are positively and significantly correlated with the trends of happiness and economic growth becomes more relevant. In the latter case the coefficient is still not significant, but becomes much bigger and the lower bound of the confidence interval is only marginally below the zero. 
 
In conclusion, the results from the long and the medium-term do not differ for SC, which is a very good predictor of SWB in both cases. Our findings differ for GDP, whose predictive capacity gives some signals of life only in the medium-term.

\clearpage
\subsection{The short-term (2 years)}
The picture depicted so far by the long and medium-term analysis -- made of null or weak correlations between the changes in SWB and GDP and robust correlations with the changes of SC -- is remarkably altered by the short-term analysis.
 
Data from the ESS allow to further reduce the length of our trends turning our attention to the relationship among biannual variations in our variables of interest.\footnote{The number of available observations is 58 short-term coefficients and not 72 as expected given the number of observations in the medium-term. Indeed, the short-term coefficients less than triple the medium-term ones because not all the countries have been surveyed in all waves. When one or more waves are missing for a given country, the number of short-term coefficients is accordingly lower. The countries for which at least one wave is missing are: Austria, Czech Republic, Estonia, Greece, Israel, Slovakia, Turkey and Ukraine.} 

\begin{figure}[htbp]
\centering
\subfloat[][\emph{Happiness and the index of social trust.\label{sp_hap_yindex23}}]
{\includegraphics[width=.65\columnwidth]{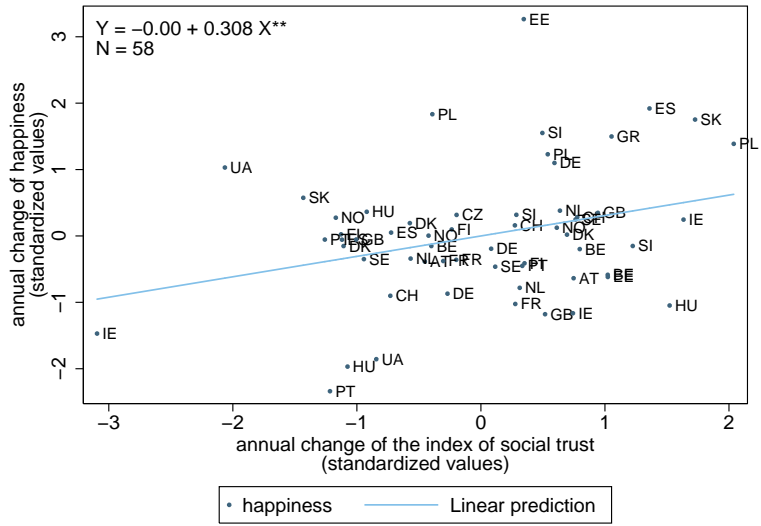}} \quad
\subfloat[][\emph{Life satisfaction and the index of social trust.\label{sp_lsat_yindex24}}]
{\includegraphics[width=.65\columnwidth]{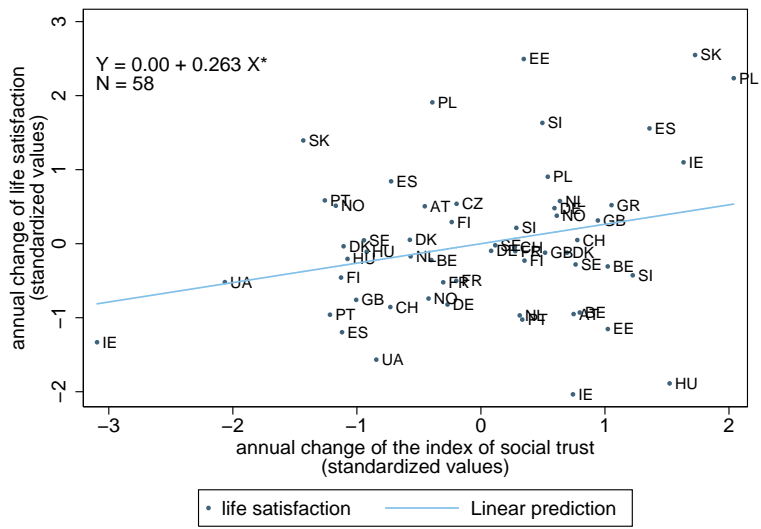}} \\
\caption{Correlations among short-term trends of subjective well-being and the changes in the index of social capital. Each dot on the scatterplots associates the short-term trend of SWB - on the y axis - with the short-term trend of the index of social trust for each country. The regression line simply depicts the correlation between the two variables.}
\label{sp_swb_factor}
\end{figure}

Figures \ref{sp_hap_yindex23} and \ref{sp_lsat_yindex24} show that in the short run both happiness and life satisfaction are positively and significantly correlated with the index of social trust. We emphasize that both the significance and the magnitude of the coefficients of social trust are much smaller and less significant than in the medium-term, in both happiness and life satisfaction regressions. 

\begin{figure}[htbp]
\centering
\subfloat[][\emph{Happiness and the the logarithm of GDP.\label{sp_hap_gdp}}]
{\includegraphics[width=.65\columnwidth]{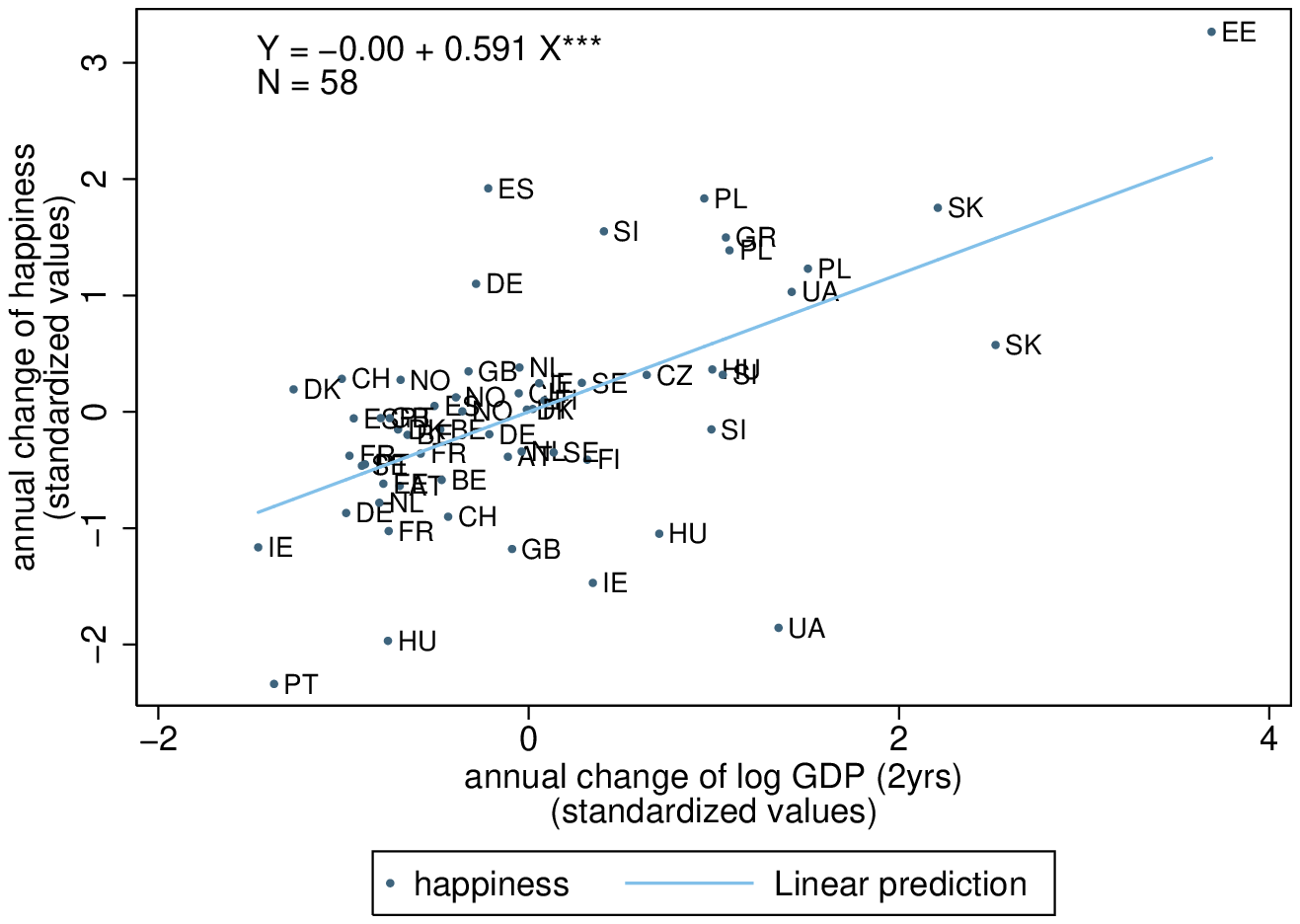}} \quad
\subfloat[][\emph{Life satisfaction and the logarithm of GDP.\label{sp_lsat_gdp}}]
{\includegraphics[width=.65\columnwidth]{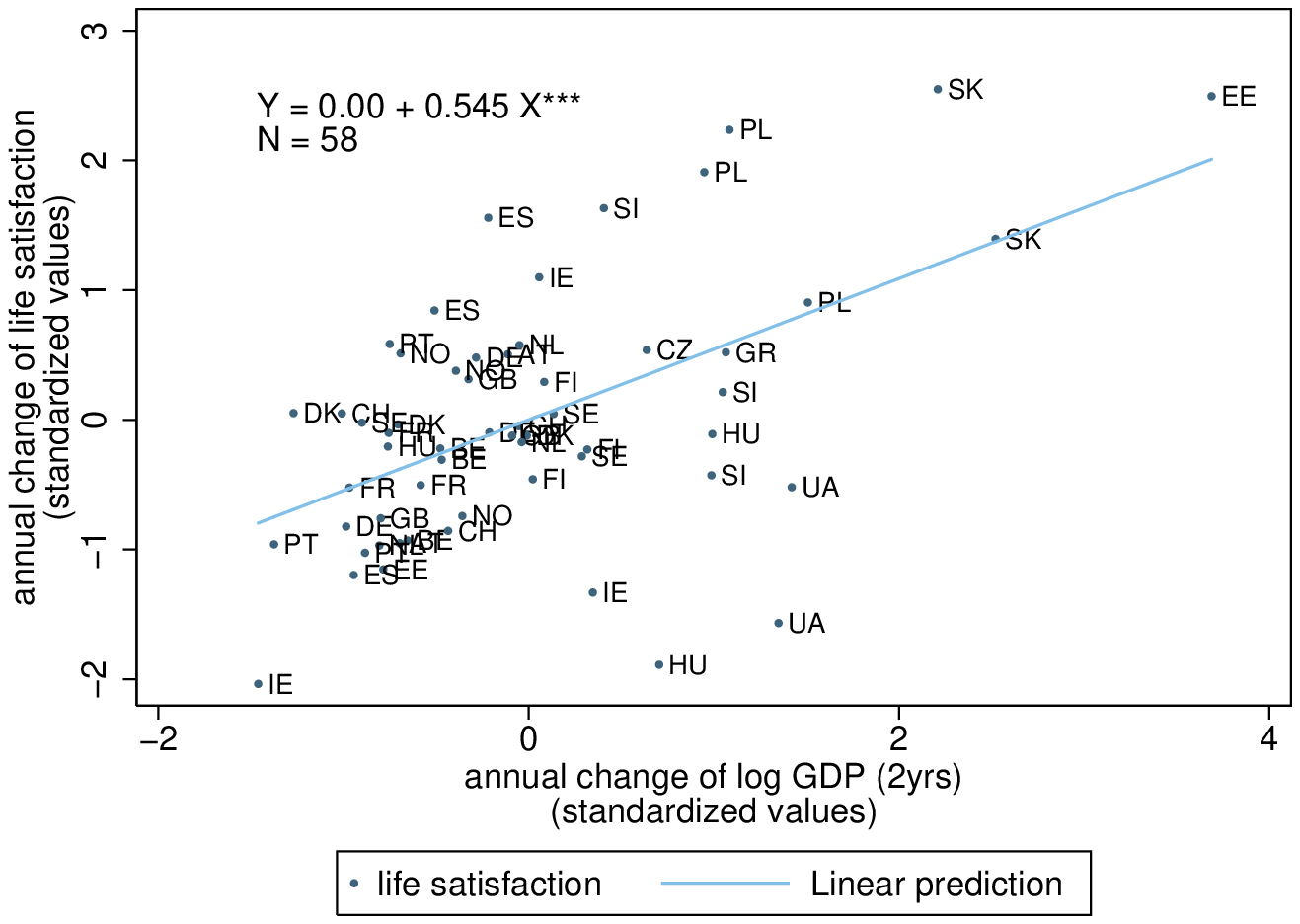}} \\
\caption{Correlations among short-term trends of subjective well-being and of the logarithm of GDP. Each dot on the scatterplots associates the short-term trend of SWB - on the y axis - with the variation of the logarithm of GDP for each country. The regression line simply depicts the correlation between the two variables.}
\label{sp_swb_factor}
\end{figure}

Figures \ref{sp_hap_gdp} and \ref{sp_lsat_gdp} provide a confirmation of previous findings in the literature. In the short run, the variations of SWB, both for happiness and life satisfaction, are largely and significantly correlated with the short-term changes of GDP. Coefficients are very large and significant at the 1\% level: an increase by one standard deviation in the variation of the logarithm of GDP is associated with more than 0.59 point increase for happiness and 0.54 point for life satisfaction (see also the second line of table \ref{sp_bivar_factor}  in  \ref{shortterm}). 

Summarizing, the bivariate analysis suggests that when considering a shorter time span, the correlation between SWB and SC remarkably weakens. Instead, the correlation between SWB and GDP sharply strengthens. 

These results are confirmed in the trivariate regressions. The first column of tab. \ref{sp_trivar_swb_factor} shows that both SC and GDP have positive and significant coefficients. Contrary to what happens when we consider longer time spans, the coefficient of the logarithm of GDP is almost 2 times larger than the SC one.  An increase by one standard deviation in the logarithm of GDP is associated with a 0.57 point increase in happiness, while the variation of social trust is associated with only 0.25 point increase. This result is even more striking when considering the second column of tab.\ref{sp_trivar_swb_factor}. Indeed, when regressing the variations of life satisfaction over the variations of SC and GDP the coefficient of social trust is not significant -- though positive -- while the coefficient of GDP is confirmed as being very large (0.52) and significant at 1\%.

\begin{table}[htbp]\centering
\def\sym#1{\ifmmode^{#1}\else\(^{#1}\)\fi}
\caption{Trivariate regressions of trends of subjective well-being over changes of the index of social trust and trends of GDP (standardized variables).}
\label{sp_trivar_swb_factor}
\begin{tabular}{l*{2}{D{.}{.}{-1}}}
\toprule
                &\multicolumn{1}{c}{(1)}&\multicolumn{1}{c}{(2)}\\
                &\multicolumn{1}{c}{happiness}&\multicolumn{1}{c}{life satisfaction}\\
\midrule
index of social trust&    0.255\sym{**} &    0.214         \\
                &   (2.18)         &   (1.58)         \\
\addlinespace
changes in log GDP (2yrs)&    0.568\sym{***}&    0.525\sym{***}\\
                &   (4.69)         &   (4.73)         \\
\addlinespace
Constant        &-3.27e-09         & 1.76e-09         \\
                &  (-0.00)         &   (0.00)         \\
\midrule
Observations    &       58         &       58         \\
Adjusted \(R^{2}\)&    0.393         &    0.318         \\
\bottomrule
\multicolumn{3}{l}{\footnotesize \textit{t} statistics in parentheses}\\
\multicolumn{3}{l}{\footnotesize \sym{*} \(p<0.10\), \sym{**} \(p<0.05\), \sym{***} \(p<0.001\)}\\
\end{tabular}
\end{table}

Summarizing, we find evidence that the changes over time of SC, as proxied by group membership and social trust, are a strong correlate of the trends of SWB. The strength of this relationship weakens when moving from the long and medium to the short-run, however. When considering trends of more than 15 years, the variations of SC are the only significant correlate of SWB, while -- as pointed out by Easterlin and colleagues -- economic growth does not play any significant predictive role. 

The size of the coefficients and their significance levels are extremely stable across models and show a remarkable pattern: moving from the medium to the short-term relationships, the coefficients of the changes of SC become about 3 times smaller. By the same token, coefficients of GDP increase by more than 2 times.
In other words, our results suggest that in the short run GDP fluctuations are closely correlated with the variation of well-being. However, this correlation is attenuated in the medium-term and wiped out in the long run.

\section{Conclusions}\label{conclu}

Available evidence documents that social capital is correlated with subjective well-being in micro data. However, the existence of a cross-sectional correlation does not imply also the existence of a  correlation between trends. Does social capital predict well-being over time as well?   The available literature exclusively focuses on the inter-temporal relationship between GDP and well-being, overlooking this issue.  The aim of our research is to compare the trends of GDP and of social capital as predictors of the trends of subjective well-being. In particular, we  run bivariate and trivariate regressions of trends of subjective well-being on trends of social capital and/or of GDP, using  a methodology for the correlation of time-series similar to the one applied by \citet{sw2008, ssw2010, ea2009} and \citet{eassz2010}.  We analyze  three different time horizons: long, medium and short-run. 

Our data sources are the World Values Survey - European Social Survey    (WVS/EVS)  for the long run  and the European Social Survey (ESS) for the medium and the short-term. These data-bases provide internationally comparable time-series on social capital and subjective well-being for many world countries (WVS/EVS) and for Europe (ESS). A major limitation of this study lies in the scarcity of time-series of social capital. The data-bases at hand allow to work only on two proxies -- although very relevant  --  of social capital: associational activity and trust. In particular, in the WVS/EVS we have to rely on the average associational activity as a proxy of social capital, while in the ESS we have to adopt  an index of social trust based on the answers to three questions about people's trustworthiness, honesty and helpfulness. In other words, the limits of the time-series on social capital do not allow us to adopt the same proxies in the long run on one side and in the medium and short run on the other. However, a check on the medium-term correlations between associational activity and SWB on the WVS/EVS confirms the medium-term findings provided in the ESS between social trust and SWB. Finally, in both data-bases, subjective well-being is proxied by happiness and life satisfaction. 

Our results suggest that the length of the time horizon of the analysis greatly matters. We find that the trends of subjective well-being over the long (15 years) and the medium-term (6 years) are largely predicted by the trends of our proxies of social capital. However, in the short-term (2 years) social capital seems to matter less. Indeed, the short run change in social trust predicts a much smaller portion of the variation of subjective well-being, compared to the medium-term. Coefficients turn out to be about 3 times smaller than in the medium run and less significant. GDP exhibits the reverse pattern compared to our proxies of social capital: %it matters more in the short than in the medium-term for the prediction of subjective well-being (OPPURE: 
its weak medium-term correlation with subjective well-being turns into a strong one in the short run. Indeed, the coefficients from   the short run regressions are almost two times larger than the ones from the medium run and more significant. Moreover, GDP is also two times more strongly correlated with SWB than social trust, while in the medium-term the correlation is more than two times weaker. As far as the long run is concerned, our findings confirm the Easterlin paradox: economic growth is unrelated to increasing well-being. 

In principle it is plausible that our results do not depend on the correlation between trends, but on random variation in the individuals surveyed driving the relationship between SWB and social capital. For instance, the sample chosen in one year in a country may be more positive than the sample chosen in another year. This positivity would be manifested in terms of their average levels of SWB, but also in terms of their average levels of social capital being higher. So there would be a common-instrument correlation. GDP, on the other hand, is being measured using a separate data source, and so does not benefit from this common-instrument correlation. As a result, the method being used is biased towards finding a stronger relationship between social capital and SWB, than between GDP and SWB.  However, there is no reason to believe that this possible bias is stronger for the long-term rather than for the medium or short-term. In  other words, this bias does not invalidate our study since it cannot affect our core results, that concern the differences between the various time spans we consider.

Summarizing, the relationship between SWB and GDP tends to vanish as time goes by. Conversely, the association between social capital and well-being seems to establish slowly and to be durable. This evidence is compatible with both the notion that income is subject to adaptation and social comparisons and with the idea that, conversely, social capital is not subject to the same forces.

\clearpage

\appendix
\section{Acknowledgments}
\acknow

\section{Descriptive tables}\label{desctabs}

\bs
\def\sym#1{\ifmmode^{#1}\else\(^{#1}\)\fi}
\caption{Availability across waves of joint observations of social capital and happiness in the WVS/EVS.}
\begin{tabular}{l*{7}{c}}
\toprule
                    &   1981-1984&   1989-1993&   1994-1999&   1999-2004&   2005-2007&   2008-2009&       Total\\
\midrule
Argentina         &        1005&        1002&        1079&        1280&        1002&           0&        5368\\
Australia         &        1228&           0&        2048&           0&        1421&           0&        4697\\
Austria           &           0&        1460&           0&        1522&           0&        1510&        4492\\
Belgium           &        1145&        2792&           0&        1912&           0&        1509&        7358\\
Brazil            &           0&        1782&        1149&           0&        1500&           0&        4431\\
Canada            &        1254&        1730&           0&        1931&        2164&           0&        7079\\
Chile             &           0&        1500&        1000&        1200&        1000&           0&        4700\\
China             &           0&        1000&        1500&        1000&        2015&           0&        5515\\
Denmark           &        1182&        1030&           0&        1023&           0&        1507&        4742\\
Finland           &        1003&         588&         987&        1038&        1014&        1134&        5764\\
France            &        1200&        1002&           0&        1615&        1001&        1501&        6319\\
Germany           &           0&        3437&        2026&        2036&        2064&        2075&       11638\\
Iceland           &         927&         702&           0&         968&           0&         808&        3405\\
Ireland           &        1217&        1000&           0&        1012&           0&        1013&        4242\\
Italy             &        1348&        2018&           0&        2000&        1012&        1519&        7897\\
Japan             &        1204&        1011&        1054&        1362&        1096&           0&        5727\\
South Korea       &           0&        1251&        1249&        1200&        1200&           0&        4900\\
Malta             &         467&         393&           0&        1002&           0&        1500&        3362\\
Mexico            &        1837&        1531&        2364&        1535&        1560&           0&        8827\\
Netherlands       &        1221&        1017&           0&        1003&        1050&        1554&        5845\\
Norway            &        1051&        1239&        1127&           0&        2115&           0&        5532\\
Portugal           &           0&        1185&           0&        1000&           0&        1553&        3738\\
South Africa      &        1596&           0&        2935&        3000&        2988&           0&       10519\\
Spain"            &        2303&        4147&        1211&        2409&        1200&        1500&       12770\\
Sweden            &         954&        1047&        1009&        1015&        1003&        1187&        6215\\
Great Britain     &        1167&        1484&           0&           0&        1041&        1561&        5253\\
United States     &        2325&        1839&        1542&        1200&        1249&           0&        8155\\
Total               &       25634&       37187&       22280&       33263&       28695&       21431&      168490\\
\midrule
Observations        &      168490&            &            &            &            &            &            \\
\bottomrule
\end{tabular}
\label{desc-a008r-01V04}
\es

\bs
\def\sym#1{\ifmmode^{#1}\else\(^{#1}\)\fi}
\caption{Availability of happiness and social capital variables across countries and waves in the ESS.}
\begin{tabular*}{15cm}{@{\hskip\tabcolsep\extracolsep\fill}l*{5}{c}}
\toprule
Countries           &  \multicolumn{4}{c}{Years}         				&       Total\\
                    &        2002&        2004&        2006&        2008&           \\
\midrule

Austria                  &        2115&        2101&        2227&           0&        6443\\
Belgium                  &        1839&        1758&        1789&        1748&        7134\\
Switzerland                  &        2013&        2115&        1783&        1795&        7706\\
Czech Republic                  &        1249&        2759&           0&        1927&        5935\\
Germany                  &        2885&        2819&        2878&        2718&       11300\\
Denmark                  &        1471&        1447&        1468&        1589&        5975\\
Estonia                  &           0&        1894&        1370&        1574&        4838\\
Spain                  &        1618&        1622&        1833&        2487&        7560\\
Finland                  &        1984&        2003&        1883&        2182&        8052\\
France                  &        1485&        1792&        1973&        2057&        7307\\
Great Britain                  &        2028&        1863&        2364&        2331&        8586\\
Greece                  &        2511&        2363&           0&        2034&        6908\\
Hungary                  &        1628&        1456&        1460&        1497&        6041\\
Ireland                  &        1945&        2216&        1717&        1756&        7634\\
Israel                  &        2352&           0&           0&        2302&        4654\\
Netherlands                  &        2336&        1868&        1870&        1760&        7834\\
Norway                  &        2032&        1748&        1741&        1540&        7061\\
Poland                  &        1991&        1626&        1631&        1520&        6768\\
Portugal                  &        1452&        1973&        2083&        2277&        7785\\
Sweden                  &        1958&        1913&        1901&        1815&        7587\\
Slovenia                  &        1453&        1390&        1419&        1241&        5503\\
Slovakia                  &           0&        1409&        1655&        1712&        4776\\
Turkey                  &           0&        1759&           0&        2115&        3874\\
Ukraina                  &           0&        1788&        1755&        1558&        5101\\
Total               &       38345&       43682&       36800&       43535&      162362\\
\midrule
Observations        &      162362&            &            &            &            \\

\bottomrule
\end{tabular*}
\label{desccountries_happy}
\es

\clearpage

\begin{small}
\begin{longtable}{lcccccc} 
\caption{Number of years between two consecutive waves of the WVS/EVS data-set when the happiness variable is available. Each columns reports the distance in years from the previous wave for each country. Figures show that the intervals vary considerably from wave to wave and from country to country. In some cases the distances are short enough to allow a short-term trend analysis, while in most cases only longer-term analysis would be possible.\label{table-a008r}}\\
\hline 
country & wave 1 & wave 2 & wave 3 & wave 4 & wave 5 & wave 6 \\ 
\hline
\endfirsthead

\multicolumn{7}{c}%
{{\tablename\ \thetable{} -- continued from previous page}} \\
\hline country & wave 1 & wave 2 & wave 3 & wave 4 & wave 5 & wave 6 \\ \hline
\endhead

\hline \multicolumn{7}{r}{\textit{continued on next page}} \\ \hline
\endfoot

\hline  
\endlastfoot

Argentina & . & 7 & 4 & 4 & 7 & . \\
Australia & . & . & 14 & . & 10 & . \\
Austria & . & . & . & 9 & . & 9 \\
Belgium & . & 9 & . & 9 & . & 10 \\
Brazil & . & . & 6 & . & 9 & . \\
Bulgaria & . & . & 7 & 2 & 7 & 2 \\
Canada & . & 8 & . & 10 & 6 & . \\
Chile & . & . & 6 & 4 & 5 & . \\
China & . & . & 5 & 6 & 6 & . \\
Czech Republic & . & . & 8 & 1 & . & 9 \\
Denmark & . & 9 & . & 9 & . & 9 \\
Estonia & . & . & 6 & 3 & . & 9 \\
Finland & . & 9 & 6 & 4 & 5 & 4 \\
France & . & 9 & . & 9 & 7 & 2 \\
Germany & . & . & 7 & 2 & 7 & 2 \\
Great Britain & . & 9 & . & . & 16 & 3 \\
Hungary & . & . & 7 & 1 & . & 9 \\
Iceland & . & 6 & . & 9 & . & 10 \\
Ireland & . & 9 & . & 9 & . & 9 \\
Italy & . & 9 & . & 9 & 6 & 4 \\
Japan & . & 9 & 5 & 5 & 5 & . \\
Latvia & . & . & 6 & 3 & . & 9 \\
Lithuania & . & . & 7 & 2 & . & 9 \\
Malta & . & 8 & . & 8 & . & 9 \\
Mexico & . & 9 & 6 & 4 & 5 & . \\
Netherlands & . & 9 & . & 9 & 7 & 2 \\
Norway & . & 8 & 6 & . & 12 & . \\
Poland & . & . & 8 & 2 & 6 & 3 \\
Portugal & . & . & . & 9 & . & 9 \\
Romania & . & . & 5 & 1 & 6 & 3 \\
Russian Federation & . & . & 5 & 4 & 7 & 2 \\
Slovakia & . & . & 8 & 1 & . & 9 \\
Slovenia & . & . & 3 & 4 & 6 & 3 \\
South Africa & . & . & 14 & 5 & 6 & . \\
South Korea & . & . & 6 & 5 & 4 & . \\
Spain & . & 9 & 5 & 4 & 8 & 1 \\
Sweden & . & 8 & 6 & 3 & 7 & 3 \\
United States & . & 8 & 5 & 4 & 7 & . \\ 
%\hline

\end{longtable}
\end{small}

\clearpage
\section{Data missingness in the WVS/EVS data-set}\label{missWVS}

Descriptive data and missing values for each variable are presented in tab.\ref{miss-a008r}  and tab.\ref{miss-a170r} for happiness and life satisfaction data, respectively. 

The numerosity of the overall sample in the two cases is substantially similar with a difference of about 1600 more observations for satisfaction with life data. Figures from the sixth column of tab.\ref{miss-a008r} and tab.\ref{miss-a170r} inform that less than 1\% of the data are missing. The only exception is represented by data about feeling of happiness. In this case the percentage of missingness  is 1.3\%. Data missingness is further explored in tab.\ref{tabmis-a008r} and tab.\ref{tabmis-a170r} where figures are contrasted over waves. In all the considered cases, percentages of missingness are of negligible size and, as such, they are not likely to bias estimates\footnote{For a more detailed discussion on data missingness and its implications for econometric analysis, please refer to \citet{schafer97, schafer1999, allison2001}}.

\begin{table}[h]
\caption{Descriptive statistics for variables jointly observed with feeling of happiness in the WVS/EVS dataset.}
\label{miss-a008r}
\resizebox{15cm}{!}{%
\begin{tabular}{lcccccc} \hline
variable & mean & sd & min & max & obs & missing \\ \hline
feeling of happiness & 3.170 & 0.674 & 1 & 4 & 166261 & 0.0132 \\
membership in at least 1 group & 0.603 & 0.489 & 0 & 1 & 167983 & 0.00301 \\
log GDP per capita & 9.503 & 0.862 & 5.970 & 10.65 & 168490 & 0 \\ \hline
\end{tabular}
}
\end{table}

\begin{table}[h]
\caption{Descriptive statistics for variables jointly observed with life satisfaction  in the WVS/EVS dataset.}
\label{miss-a170r}
\resizebox{15cm}{!}{%
\begin{tabular}{lcccccc} \hline
variable & mean & sd & min & max & obs & missing \\ \hline
satisfaction with life & 7.355 & 2.068 & 1 & 10 & 167906 & 0.00771 \\
membership in at least 1 group & 0.598 & 0.490 & 0 & 1 & 168704 & 0.00300 \\
log GDP per capita & 9.501 & 0.867 & 5.970 & 10.65 & 169211 & 0 \\ \hline
\end{tabular}
}
\end{table}

\begin{table}[h]
\caption{Percentage of data missingness across waves for variables jointly observed with feeling of happiness  in the WVS/EVS dataset.}
\label{tabmis-a008r}
\resizebox{15cm}{!}{%
\begin{tabular}{lccccccc} \hline
variable & wave 1 & wave 2 & wave 3 & wave 4 & wave 5 & wave 6 & total \\ \hline
feeling of happiness & 0.0209 & 0.0256 & 0.00678 & 0.00776 & 0.00634 & 0.00705 & 166261 \\
membership in at least 1 group & 0 & 0 & 0.00148 & 0 & 0.00227 & 0.0191 & 167983 \\
log GDP per capita & 0 & 0 & 0 & 0 & 0 & 0 & 168490 \\ \hline
\end{tabular}
}
\end{table}

\begin{table}[h]
\caption{Percentage of data missingness across waves  for variables jointly observed with life satisfaction in the WVS/EVS dataset.}
\label{tabmis-a170r}
\resizebox{15cm}{!}{%
\begin{tabular}{lccccccc} 
\toprule
variable & wave 1 & wave 2 & wave 3 & wave 4 & wave 5 & wave 6 & total \\ \hline
satisfaction with life & 0.0122 & 0.00764 & 0.00571 & 0.00785 & 0.00746 & 0.00434 & 167906 \\
membership in at least 1 group & 0 & 0 & 0.00157 & 0 & 0.00227 & 0.0191 & 168704 \\
log GDP per capita & 0 & 0 & 0 & 0 & 0 & 0 & 169211 \\ \hline
\end{tabular}
}
\end{table}

\clearpage
\section{Data missingness in the ESS data-set}
\label{missessapp}
The sixth column of tab.\ref{miss-ESS} informs that the percentage of missing data is on average less than 1\%. Only in the case of the index of social trust the percentage of missingness raises to 1.4\%. However, such a small percentage does not raise any particular worry for the reliability of our estimates \citep{allison2001}. Data missingness is further analysed across waves in tab.\ref{tabmiss-ESS}. Figures inform that also in this case percentages of missingness are negligible and, according to the literature on data missingness, they are not likely to affect estimates \citep{schafer97, schafer1999, allison2001}.

\begin{table}[htbp]
\caption{Descriptive statistics for variables in the ESS data-set}
\label{miss-ESS}
\resizebox{15cm}{!}{%
\begin{tabular}{lcccccc} \hline
variable & mean & sd & min & max & obs & missing \\ \hline
How happy are you & 7.231 & 2.007 & 0 & 10 & 167190 & 0.00663 \\
How satisfied with life as a whole & 6.875 & 2.319 & 0 & 10 & 167209 & 0.00652 \\
Most people try to take advantage of you, or try t & 5.567 & 2.378 & 0 & 10 & 166676 & 0.00968 \\
Most people can be trusted or you can't be too car & 4.982 & 2.493 & 0 & 10 & 167571 & 0.00437 \\
Most of the time people helpful or mostly looking  & 4.801 & 2.385 & 0 & 10 & 167283 & 0.00608 \\
index of social trust & -4.38e-10 & 1 & -2.553 & 2.424 & 165760 & 0.0151 \\
 log GDP per capita & 9.673 & 0.819 & 6.613 & 10.93 & 168306 & 0 \\ \hline
\end{tabular}
}
\end{table}

\begin{table}[htbp]
\caption{Percentage of data missingness across waves in the ESS data-set.}
\label{tabmiss-ESS}
\resizebox{15cm}{!}{%
\begin{tabular}{lccccc} \hline
variable & wave 1 & wave 2 & wave 3 & wave 4 & total \\ \hline
How happy are you & 0.00515 & 0.00600 & 0.00752 & 0.00781 & 167190 \\
How satisfied with life as a whole & 0.00626 & 0.00490 & 0.00608 & 0.00874 & 167209 \\
Most people try to take advantage of you, or try t & 0.00959 & 0.0101 & 0.0108 & 0.00839 & 166676 \\
Most people can be trusted or you can't be too car & 0.00454 & 0.00441 & 0.00568 & 0.00305 & 167571 \\
Most of the time people helpful or mostly looking  & 0.00613 & 0.00638 & 0.00676 & 0.00516 & 167283 \\
index of social trust & 0.0152 & 0.0160 & 0.0168 & 0.0127 & 165760 \\
 log GDP per capita & 0 & 0 & 0 & 0 & 168306 \\ \hline
\end{tabular}
}
\end{table}

\clearpage
\section{Factor analysis for trust questions in the ESS}
\label{facapp}
Tab. \ref{factor_tot} informs that in the pooled sample, factor loadings range from .80 to .85 thus suggesting that the three variables contribute equally to the definition of a latent concept that we call ``social trust''. When observing results across waves (see tab.\ref{factor}), we notice that discrepancies arise mainly in the first and third wave where factor loadings range from about .79 for the helpfulness variable to .84 for the fairness variable. The slight variability among factor loadings both in the pooled sample and within waves convinced us of the opportunity to build an aggregated index of social trust resulting from the standardized weighted average of the three items.

\begin{table}[htbp]\centering
\def\sym#1{\ifmmode^{#1}\else\(^{#1}\)\fi}
\caption{Factor loading and unique variances for the pooled sample}
\label{factor_tot}
\begin{tabular}{l*{3}{D{.}{.}{-1}}}
\toprule
                    &    Factor 1&         Psi\\
\midrule
Most people try to take advantage of you&    .850&    .276\\
Most people can be trusted &    .840&    .293\\
Most of the time people helpful &    .804&    .352\\
\bottomrule
\end{tabular}
\end{table}

\begin{table}[htbp]\centering
\caption{Factor loading and unique variances across waves}
\label{factor}
\begin{tabular}{l*{3}{D{.}{.}{-1}}}
\toprule
wave 1                & Factor 1 &    Psi   \\
						& 		&			\\

Most people try to take advantage of you&    .849 &    .278\\
Most people can be trusted &    .836&    .299 \\
Most of the time people helpful &    .794 &    .368\\

\midrule

wave 2                  &  &      \\
						& 		&			\\
Most people try to take advantage of you&    .844&     .287\\
Most people can be trusted &    .835&    .301\\
Most of the time people helpful &    .804&    .353\\

\midrule

wave 3                    &  &     \\
						& 		&			\\
						
Most people try to take advantage of you&    .844&    .286\\
Most people can be trusted &    .834&    .304\\
Most of the time people helpful &    .797&     .364\\
	
\midrule

wave 4                   &  &    \\
						& 		&			\\

Most people try to take advantage of you&    .861&    .257\\
Most people can be trusted&    .852&    .272\\
Most of the time people helpful &    .818&    .330\\

\bottomrule
\end{tabular}
\end{table}

\clearpage
\section{Medium-term relationships deleting Turkey}
\label{mediumterm}
%\begin{table}[htbp]\centering
%\def\sym#1{\ifmmode^{#1}\else\(^{#1}\)\fi}
%\caption{Bivariate correlations among medium-term trends of subjective well-being and changes of the index of social trust excluding Turkey (standardized variables).}
%\label{lp_bivar_swb_factor-turkey}
%\begin{tabular}{l*{4}{D{.}{.}{-1}}}
%\toprule
%                &\multicolumn{1}{c}{(1)}&\multicolumn{1}{c}{(2)}&\multicolumn{1}{c}{(3)}&\multicolumn{1}{c}{(4)}\\
%                &\multicolumn{1}{c}{happiness}&\multicolumn{1}{c}{happiness}&\multicolumn{1}{c}{life satisfaction}&\multicolumn{1}{c}{life satisfaction}\\
%\midrule
%index of social trust&    0.541\sym{**} &                  &    0.725\sym{***}&                  \\
%                &   (2.81)         &                  &   (4.24)         &                  \\
%\addlinespace
%trend of log GDP&                  &    0.390\sym{**} &                  &    0.420\sym{**} \\
%                &                  &   (3.47)         &                  &   (2.23)         \\
%\addlinespace
%Constant        &   0.0783         &    0.164         &  0.00652         &    0.119         \\
%                &   (0.67)         &   (1.44)         &   (0.05)         &   (0.74)         \\
%\midrule
%Observations    &       23         &       23         &       23         &       23         \\
%
%\bottomrule
%\multicolumn{5}{l}{\footnotesize \textit{t} statistics in parentheses}\\
%\multicolumn{5}{l}{\footnotesize \sym{*} \(p<0.10\), \sym{**} \(p<0.05\), \sym{***} \(p<0.001\)}\\
%\end{tabular}
%\end{table}

\begin{table}[htbp]\centering
\def\sym#1{\ifmmode^{#1}\else\(^{#1}\)\fi}
\caption{Trivariate regressions of medium-term trends of subjective well-being over changes of the index of social trust and trends of GDP excluding Turkey (standardized variables).}
\label{lp_trivar_swb_factor-turkey}
\begin{tabular}{l*{2}{D{.}{.}{-1}}}
\toprule
                &\multicolumn{1}{c}{happiness}&\multicolumn{1}{c}{life satisfaction}\\
\midrule
index of social trust&    0.461\sym{***}&    0.643\sym{***}\\
                &   (5.80)         &   (4.21)         \\
\addlinespace
trend of log GDP&    0.333\sym{***}&    0.341\sym{**} \\
                &   (5.09)         &   (2.14)         \\
\addlinespace
Constant        &   0.0968         &   0.0255         \\
                &   (1.10)         &   (0.20)         \\
\midrule
Observations    &       23         &       23         \\
Adjusted \(R^{2}\)&    0.572         &    0.487         \\
\bottomrule
\multicolumn{3}{l}{\footnotesize \textit{t} statistics in parentheses}\\
\multicolumn{3}{l}{\footnotesize \sym{*} \(p<0.10\), \sym{**} \(p<0.05\), \sym{***} \(p<0.001\)}\\
\end{tabular}
\end{table}

\begin{table}[htbp]\centering
\def\sym#1{\ifmmode^{#1}\else\(^{#1}\)\fi}
\caption{Trivariate regressions of trends of subjective well-being over changes of the index of social trust and trends of GDP excluding Turkey and transition economies (standardized values).}
\label{lp_trivar_swb_factor-tt}
\begin{tabular}{l*{2}{D{.}{.}{-1}}}
\toprule
                &\multicolumn{1}{c}{happiness}&\multicolumn{1}{c}{life satisfaction}\\
\midrule
\addlinespace
index of social trust&    0.419\sym{*}  &    0.628\sym{*}  \\
                &   (1.84)         &   (2.01)         \\
\addlinespace
trend of log GDP&    0.314         &  -0.0529         \\
                &   (1.16)         &  (-0.14)         \\
\addlinespace
Constant        &    0.137         &   -0.151         \\
                &   (0.71)         &  (-0.57)         \\
\midrule
Observations    &       16         &       16         \\
Adjusted \(R^{2}\)&    0.241         &    0.133         \\
\bottomrule
\multicolumn{3}{l}{\footnotesize \textit{t} statistics in parentheses}\\
\multicolumn{3}{l}{\footnotesize \sym{*} \(p<0.10\), \sym{**} \(p<0.05\), \sym{***} \(p<0.001\)}\\
\end{tabular}
\end{table}

\clearpage
\section{Short-term relationships}
\label{shortterm}
\begin{table}[htbp]\centering
\def\sym#1{\ifmmode^{#1}\else\(^{#1}\)\fi}
\caption{Bivariate correlations among short-term trends of subjective well-being and changes of the index of social trust (standardized variables).}
\label{sp_bivar_factor}
\resizebox{15cm}{!}{%
\begin{tabular}{l*{4}{D{.}{.}{-1}}}
\toprule
                &\multicolumn{1}{c}{(1)}&\multicolumn{1}{c}{(2)}&\multicolumn{1}{c}{(3)}&\multicolumn{1}{c}{(4)}\\
                &\multicolumn{1}{c}{happiness}&\multicolumn{1}{c}{happiness}&\multicolumn{1}{c}{life satisfaction}&\multicolumn{1}{c}{life satisfaction}\\
\midrule
index of social trust&    0.308\sym{**} &                  &    0.263\sym{*}  &                  \\
                &   (2.33)         &                  &   (1.80)         &                  \\
\addlinespace
changes in log GDP (2yrs)&                  &    0.591\sym{***}&                  &    0.545\sym{***}\\
                &                  &   (4.65)         &                  &   (4.66)         \\
\addlinespace
Constant        &-2.54e-09         &-4.92e-09         & 2.42e-09         & 3.71e-10         \\
                &  (-0.00)         &  (-0.00)         &   (0.00)         &   (0.00)         \\
\midrule
Observations    &       58         &       58         &       58         &       58         \\
\bottomrule
\multicolumn{5}{l}{\footnotesize \textit{t} statistics in parentheses}\\
\multicolumn{5}{l}{\footnotesize \sym{*} \(p<0.10\), \sym{**} \(p<0.05\), \sym{***} \(p<0.001\)}\\
\end{tabular}}
\end{table}

\clearpage
\section{Estimates excluding countries with sampling problems}
\label{probsampling}
\begin{table}[htbp]\centering
\def\sym#1{\ifmmode^{#1}\else\(^{#1}\)\fi}
\caption{Bivariate correlations among long-term trends of happiness with SC and GDP respectively (standardized variables).}
\begin{tabular}{l*{2}{D{.}{.}{-1}}}
\toprule
                &\multicolumn{1}{c}{happiness}         &\multicolumn{1}{c}{happiness}         \\
\midrule
membership in group or association &    0.756\sym{**} &                  \\
                &   (3.06)         &                  \\
\addlinespace
log GDP&                  &   -0.160         \\
                &                  &  (-0.80)         \\
\addlinespace
Constant        &   -0.697\sym{**} &   -0.386\sym{**} \\
                &  (-3.58)         &  (-2.13)         \\
\midrule
Observations    &       23         &       23         \\
\bottomrule
\multicolumn{3}{l}{\footnotesize \textit{t} statistics in parentheses}\\
\multicolumn{3}{l}{\footnotesize \sym{*} \(p<0.10\), \sym{**} \(p<0.05\), \sym{***} \(p<0.001\)}\\
\end{tabular}
\end{table}

\begin{table}[htbp]\centering
\def\sym#1{\ifmmode^{#1}\else\(^{#1}\)\fi}
\caption{Bivariate correlations among long-term trends of life satisfaction with SC and GDP respectively (standardized variables).}
\begin{tabular}{l*{2}{D{.}{.}{-1}}}
\toprule
                &\multicolumn{1}{c}{life satisfaction}         &\multicolumn{1}{c}{life satisfaction}         \\
\midrule
membership in group or association &    0.353\sym{***}&                  \\
                &   (4.35)         &                  \\
\addlinespace
log GDP&                  &   0.0738         \\
                &                  &   (0.64)         \\
\addlinespace
Constant        &   -0.723\sym{***}&   -0.544\sym{***}\\
                &  (-8.11)         &  (-4.63)         \\
\midrule
Observations    &       23         &       23         \\
\bottomrule
\multicolumn{3}{l}{\footnotesize \textit{t} statistics in parentheses}\\
\multicolumn{3}{l}{\footnotesize \sym{*} \(p<0.10\), \sym{**} \(p<0.05\), \sym{***} \(p<0.001\)}\\
\end{tabular}
\end{table}

\begin{table}[htbp]\centering
\def\sym#1{\ifmmode^{#1}\else\(^{#1}\)\fi}
\caption{Trivariate regressions of long-term trends of happiness over trends of SC and GDP (standardized variables).}
\begin{tabular}{l*{1}{D{.}{.}{-1}}}
\toprule
                &\multicolumn{1}{c}{happiness}\\
\midrule
membership in group or association &    0.755\sym{**} \\
                &   (2.98)         \\
\addlinespace
log GDP& -0.00176         \\
                &  (-0.01)         \\
\addlinespace
Constant        &   -0.697\sym{**} \\
                &  (-3.49)         \\
\midrule
Observations    &       23         \\
Adjusted \(R^{2}\)&    0.309         \\
\bottomrule
\multicolumn{2}{l}{\footnotesize \textit{t} statistics in parentheses}\\
\multicolumn{2}{l}{\footnotesize \sym{*} \(p<0.10\), \sym{**} \(p<0.05\), \sym{***} \(p<0.001\)}\\
\end{tabular}
\end{table}

\begin{table}[htbp]\centering
\def\sym#1{\ifmmode^{#1}\else\(^{#1}\)\fi}
\caption{Trivariate regressions of long-term trends of life satisfaction over trends of SC and GDP (standardized variables).}
\begin{tabular}{l*{1}{D{.}{.}{-1}}}
\toprule
                &\multicolumn{1}{c}{life satisfaction}\\
\midrule
membership in group or association &    0.362\sym{***}\\
                &   (4.13)         \\
\addlinespace
log GDP&   0.0908         \\
                &   (1.25)         \\
\addlinespace
Constant        &   -0.711\sym{***}\\
                &  (-7.77)         \\
\midrule
Observations    &       23         \\
Adjusted \(R^{2}\)&    0.214         \\
\bottomrule
\multicolumn{2}{l}{\footnotesize \textit{t} statistics in parentheses}\\
\multicolumn{2}{l}{\footnotesize \sym{*} \(p<0.10\), \sym{**} \(p<0.05\), \sym{***} \(p<0.001\)}\\
\end{tabular}
\end{table}

\clearpage
\section{List of groups and associations mentioned in the WVS/EVS questionnaire}
\label{list}

Respondents were asked to mention whether they belonged or were performing unpaid voluntary work for any of the following list of associations:
\begin{itemize}
\item{social welfare service for elderly;}
\item{religious organization;}
\item{education, arts, music or cultural activities;}
\item{labour unions;}
\item{political parties;}
\item{local political actions;}
\item{human rights;}
\item{conservation, the environment, ecology, animal rights;}
\item{conservation, the environment, ecology;}
\item{animal rights;}
\item{professional associations;}
\item{youth work;}
\item{sports or recreation;}
\item{women's group;}
\item{peace movement;}
\item{organization concerned with health;}
\item{consumer groups;}
\item{other groups.}
\end{itemize}

\clearpage
\section{Country acronyms in the ESS}
\label{acronyms}

\begin{tabular}{lm{3cm}l}
AT: Austria & & HU: Hungary \\
BE: Belgium & & IE: Ireland \\
CH: Switzerland  & & IL: Israel \\
CZ: Czech Republic & & NL: Netherlands \\
DE: Germany & & NO: Norway \\
DK: Denmark & &  PL: Poland \\
EE: Estonia & & PT: Portugal \\
ES: Spain & & SE: Sweden \\
FI: Finland & & SI: Slovenia \\
FR: France & & SK: Slovakia \\
GB: Great Britain & & TR: Turkey \\
GR: Greece & &  UA: Ukraina 
\end{tabular}

\clearpage

%\bibliographystyle{model2-names}
%\bibliography{library}

\end{document}